\documentclass[prl,twocolumn,longbibliography,amsmath,superscriptaddress,amssymb,nobibnotes]{revtex4-1}

\usepackage{graphicx}
\usepackage{dcolumn}
\usepackage{bm}
\usepackage{amssymb}
\usepackage{amsmath}
\usepackage{color}
\usepackage[dvipsnames]{xcolor}
\usepackage{placeins}
\usepackage{epstopdf}

\usepackage[normalem]{ulem}

\usepackage[linkcolor=blue,urlcolor=blue,citecolor=blue,colorlinks=true]{hyperref}

\newcommand{\rtm}[1]{\mathrm{#1}}

\begin{document}

\title{Observation of a cubic Rashba effect in the surface spin structure of rare-earth ternary materials}

\author{D.~Yu.~Usachov}
\affiliation{St. Petersburg State University, 7/9 Universitetskaya nab., St. Petersburg, 199034, Russia}

\author{I.~A.~Nechaev}
\affiliation{Department of Electricity and Electronics, FCT-ZTF, UPV-EHU, 48080 Bilbao, Spain}

\author{G. Poelchen}
\affiliation{Institut f\"{u}r Festk\"{o}rperphysik, Technische Universit\"{a}t Dresden, D-01062 Dresden, Germany}

\author{M.~G\"{u}ttler}
\affiliation{Institut f\"{u}r Festk\"{o}rperphysik, Technische Universit\"{a}t Dresden, D-01062 Dresden, Germany}

\author{E.~E.~Krasovskii}
\affiliation{Donostia International Physics Center (DIPC), 20018 Donostia/San Sebasti\'{a}n, Basque Country, Spain}
\affiliation{Departamento de F\'{\i}sica de Materiales UPV/EHU, 20080 Donostia/San Sebasti\'{a}n, Basque Country, Spain}
\affiliation{IKERBASQUE, Basque Foundation for Science, 48013, Bilbao, Spain}

\author{S.~Schulz}
\affiliation{Institut f\"{u}r Festk\"{o}rperphysik, Technische Universit\"{a}t Dresden, D-01062 Dresden, Germany}

\author{A.~Generalov}
\affiliation{Max IV Laboratory, Lund University, Box 118, 22100 Lund, Sweeden}

\author{K.~Kliemt}
\affiliation{Kristall- und Materiallabor, Physikalisches Institut, Goethe-Universit\"{a}t Frankfurt, Max-von-Laue Strasse 1, D-60438 Frankfurt am Main, Germany}

\author{A.~Kraiker}
\affiliation{Kristall- und Materiallabor, Physikalisches Institut, Goethe-Universit\"{a}t Frankfurt, Max-von-Laue Strasse 1, D-60438 Frankfurt am Main, Germany}

\author{C.~Krellner}
\affiliation{Kristall- und Materiallabor, Physikalisches Institut, Goethe-Universit\"{a}t Frankfurt, Max-von-Laue Strasse 1, D-60438 Frankfurt am Main, Germany}

\author{K. Kummer}
\affiliation{European Synchrotron Radiation Facility, 71 Avenue des Martyrs, Grenoble, France}

\author{S.~Danzenb\"{a}cher}
\affiliation{Institut f\"{u}r Festk\"{o}rperphysik, Technische Universit\"{a}t Dresden, D-01062 Dresden, Germany}

\author{C.~Laubschat}
\affiliation{Institut f\"{u}r Festk\"{o}rperphysik, Technische Universit\"{a}t Dresden, D-01062 Dresden, Germany}

\author{A.~P.~Weber}
\affiliation{Donostia International Physics Center (DIPC), 20018 Donostia/San Sebasti\'{a}n, Basque Country, Spain}

\author{E.~V.~Chulkov}
\affiliation{St. Petersburg State University, 7/9 Universitetskaya nab., St. Petersburg, 199034, Russia}
\affiliation{Donostia International Physics Center (DIPC), 20018 Donostia/San Sebasti\'{a}n, Basque Country, Spain}
\affiliation{Departamento de F\'{\i}sica de Materiales UPV/EHU, 20080 Donostia/San Sebasti\'{a}n, Basque Country, Spain}
\affiliation{Centro de F\'{\i}sica de Materiales CFM-MPC and Centro Mixto CSIC-UPV/EHU, 20018 Donostia/San Sebasti\'{a}n, Basque Country, Spain}
\affiliation{Tomsk State University, Lenina Av. 36, 634050, Tomsk, Russia}

\author{A.~F.~Santander-Syro}
\affiliation{Universit\'e Paris-Saclay, CNRS,  Institut des Sciences Mol\'eculaires d'Orsay, 91405, Orsay, France}

\author{T.~Imai}
\affiliation{Graduate School of Science, Hiroshima University, 1-3-1 Kagamiyama, Higashi-Hiroshima 739-8526, Japan}

\author{K.~Miyamoto}
\affiliation{Hiroshima Synchrotron Radiation Center, Hiroshima University, 2-313 Kagamiyama, Higashi-Hiroshima 739-0046, Japan}

\author{T.~Okuda}
\affiliation{Hiroshima Synchrotron Radiation Center, Hiroshima University, 2-313 Kagamiyama, Higashi-Hiroshima 739-0046, Japan}

\author{D.~V.~Vyalikh}
\affiliation{Donostia International Physics Center (DIPC), 20018 Donostia/San Sebasti\'{a}n, Basque Country, Spain}
\affiliation{IKERBASQUE, Basque Foundation for Science, 48013, Bilbao, Spain}

\begin{abstract}
Spin-orbit interaction and structure inversion asymmetry in combination with magnetic ordering is a promising route to novel materials with highly mobile spin-polarized carriers at the surface. Spin-resolved measurements of the photoemission current from the Si-terminated surface of the antiferromagnet TbRh$_2$Si$_2$ and their analysis within an {\em ab initio } one-step theory unveil an unusual triple winding of the electron spin along the fourfold-symmetric constant energy contours of the surface states. A two-band ${\mathbf k}\cdot{\mathbf p}$ model is presented that yields the triple winding as a cubic Rashba effect. The curious in-plane spin-momentum locking is remarkably robust and remains intact across a paramagnetic-antiferromagnetic transition in spite of spin-orbit interaction on Rh atoms being considerably weaker than the out-of-plane exchange field due to the Tb 4$f$ moments.
\end{abstract}

\maketitle
	
%%%%%%%%%%%%%%%%%%%%%%%%%%%%
\begin{figure*}[t]
\includegraphics[width=\linewidth]{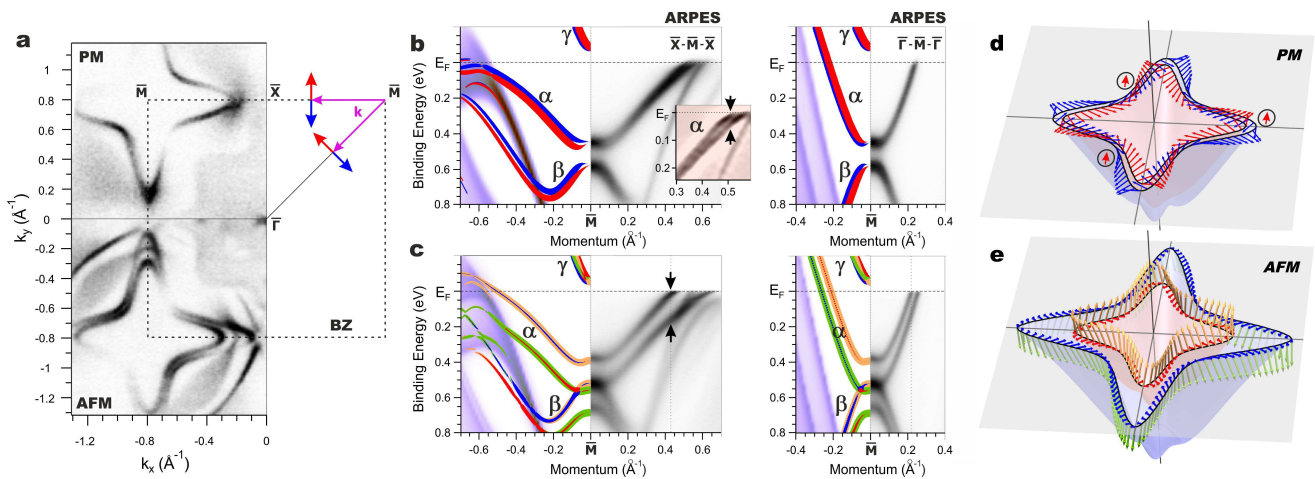}
\caption{(a) Fermi contours from ARPES for PM and AFM phases. Red and blue arrows indicate color schemes used for in-plane spin components derived from DFT calculation in graphs (b) and (c). Calculated and measured band structure for Si-terminated TbRh$_2$Si$_2$ for PM (b) and AFM (c) phases. The inset in (b) shows the spin-orbit splitting of the $\alpha$ state: black arrows indicate the 35~meV splitting. Black arrows in (c) indicate the 140~meV splitting of the $\alpha$ state in AFM phase.
Violet-brown palette shows calculated surface-projected bulk states. For the surface states the in-plane spin polarization is shown in red and blue, while yellow and green highlight the out-of-plane positive and negative spin component $S_z$, respectively.
Spin resolved {\em ab initio } CECs for the state $\alpha$ in the PM (d) and AFM (e) phase calculated at the binding energy of 0.23~eV. The in-plane spin orientation, $\mathrm{\mathbf{S}}_{\parallel}$, is indicated by red (blue) arrows for the inner (outer) contours. In AFM phase, the color of the spin vector $\mathrm{\mathbf{S}}= \mathrm{\mathbf{S}}_{\parallel}+S_z\hat{\mathrm{\mathbf{z}}}$ shows the sign of the out-of-plane spin  $S_z$ as in Fig.~\ref{fig:1}(c).}
\label{fig:1}
\end{figure*}
%%%%%%%%%%%%%%%%%%%%%%%%%%%%

Spin-orbit interaction (SOI) and its combination with exchange magnetic~\cite{Susanne2019} or Kondo interactions~\cite{Generalov2018} in non-centrosymmetric two-dimensional (2D) systems provides a versatile basis for the design of novel materials for magnetic applications. Combination of these fundamental interactions allows one to create spin-polarized carriers trapped at the surface and to fine-tune their properties. The underlying phenomenon is the splitting of spin-degenerate 2D states: a momentum-dependent spin-orbit splitting governed by the Rashba effect and a Zeeman-like splitting induced by the magnetic exchange interaction. The simultaneous action of these mechanisms can bring about a puzzling momentum dependence of the spin polarization and an exotic spin structure at the Fermi contour very different from textbook  examples~\cite{Generalov2018, Nechaev_PRB_2018, Susanne2019, Nechaev_PRB_2019}.

The Rashba effect turns out to significantly influence the spin properties of the carriers in magnetic systems even when the exchange interaction is much stronger than SOI~\cite{Nechaev_PRB_2018}. For this reason, the main emphasis has been on the SOI-induced energy splitting and on in-plane spin-momentum locking, which often deviates from the prediction of the classical Rashba model with its helical effective magnetic field (EMF): In reality, the ${\mathbf k}$-dependence of the splitting is far from linear, and the spin-momentum locking is not orthogonal~\cite{Bondarenko_SciRep_2013, Friedrich_NJP_2017}.

Vivid examples are found in semiconductor quantum wells~\cite{MoriyaPRL_2014, Hong_PRL_2018} and oxide surfaces and interfaces~\cite{Nakamura_PRL_2012, King_NatComm_2014, Varignon_NatPhys_2018, Gariglio_RPP_2018, WeinanLin2019}, which prominently feature the so-called cubic Rashba effect responsible for the non-linear ($\propto|{\mathbf k}|^3$) dependence of the spin-orbit splitting of 2D heavy-hole states~\cite{Gerchikov_SPS_1992, Winkler_KP, Marcellina_PRB_2017}. Remarkably, the EMF in the cubic Rashba model has a significantly different symmetry from the linear model, so that the in-plane {\it pseudospin} of the heavy hole rotates three times faster in moving around the Fermi contour and is not anymore orthogonal to ${\mathbf k}$ everywhere~\cite{Nakamura_PRL_2012, Wang_PRB_2012, Kondo_NJP_2015, Bladwell_PRB_2015, Shanavas_PRB_2016}.

In this work, using spin- and angle-resolved photoelectron spectroscopy (SR-ARPES) we unveil a system that realizes a cubic Rashba effect for the {\em true spin}: the Si-terminated surface of the antiferromagnetic (AFM) TbRh$_2$Si$_2$ (TRS). It belongs to a family of RET$_2$Si$_2$ materials (RE and T are rare-earth and transition metal atoms, respectively) of the ThCr$_2$Si$_2$ type~\cite{Felner_JPCS_1984}. Here, we report the first observation of the exotic spin structure predicted {\em ab initio } for the RET$_2$Si$_2$ family~\cite{Nechaev_PRB_2018}, which has the distinctive triple winding of the in-plane spin~\cite{Susanne2019} along the fourfold-symmetric constant energy contour (CEC). In the present context, the term {\em spin} refers to the expectation value of the spin operator rather than to a {\em pseudospin}. We corroborate our observation with a calculation of the spin photocurrent within an {\em ab initio } one-step theory~\cite{Kimura2010,Bentmann2017}. Furthermore, we derive from the full microscopic Hamiltonian a minimal relativistic ${\mathbf k}\cdot{\mathbf p}$ model that proves the observed spin structure to be due to the cubic Rashba effect.

The ARPES experiments were performed at the I05 beamline of the Diamond Light Source, while the SR-ARPES measurements were conducted at the ESPRESSO instrument (BL-9B) of the HiSOR facility \cite{Okuda_RSI_2011} and at SR-ARPES instrument of RGBL-2 beamline at BESSY-II. These instruments are equipped with VLEED- and Mott- detectors for spin analysis, respectively. All presented data were taken using a photon energy of 55~eV and linear polarization. Single-crystalline samples of TbRh$_2$Si$_2$ were grown using a high-temperature indium-flux method~\cite{Kliemt_2019}. The calculations in Fig.~\ref{fig:1} were performed with the full-potential local orbital method~\cite{Koepernik1999} within density functional theory (DFT), see Supplemental Material (SM)~\cite{Note1}. The ARPES spectra were calculated within the one-step theory~\cite{Adawi64}, in which the outgoing electron is described by the time-reversed LEED state. It is calculated scalar-relativistically as explained in Ref.~\cite{Krasovskii99}. The initial states are eigenfunctions of a thick slab obtained from a large-size relativistic ${\mathbf k}\cdot{\mathbf p}$ Hamiltonian, see the SM~\cite{Note1}.

Antiferromagnetic TbRh$_2$Si$_2$ has the N\'eel temperature of $\sim$90~K. In the AFM phase the Tb $4f$ moments in the {\it ab}-plane are ferromagnetically (FM) ordered with out-of-plane orientation~\cite{Quezel_TRS_1984}. The neighboring planes of Tb are separated by silicide Si--Rh--Si blocks and the ordered Tb 4$f$ moments couple antiferromagnetically along the {\it c} axis. Upon cleavage, the resulting surface exhibits either Tb or Si termination. The latter has surface states in a large projected band gap around the $\bar{M}$-point. Figure~\ref{fig:1}(a) shows Fermi contours  in PM and AFM phases. The surface states are seen as four-point stars around the corners of the surface Brillouin zone (SBZ) with a strong splitting typical of the AFM-ordered Si-terminated RET$_2$Si$_2$ materials~\cite{Generalov2018, Chick14, YbCoSi_2, GdRhSi, HoRhSi}.

%%%%%%%%%%%%%%%%%%%%%%%%%%%%
\begin{figure}[t]
\includegraphics[width=\linewidth]{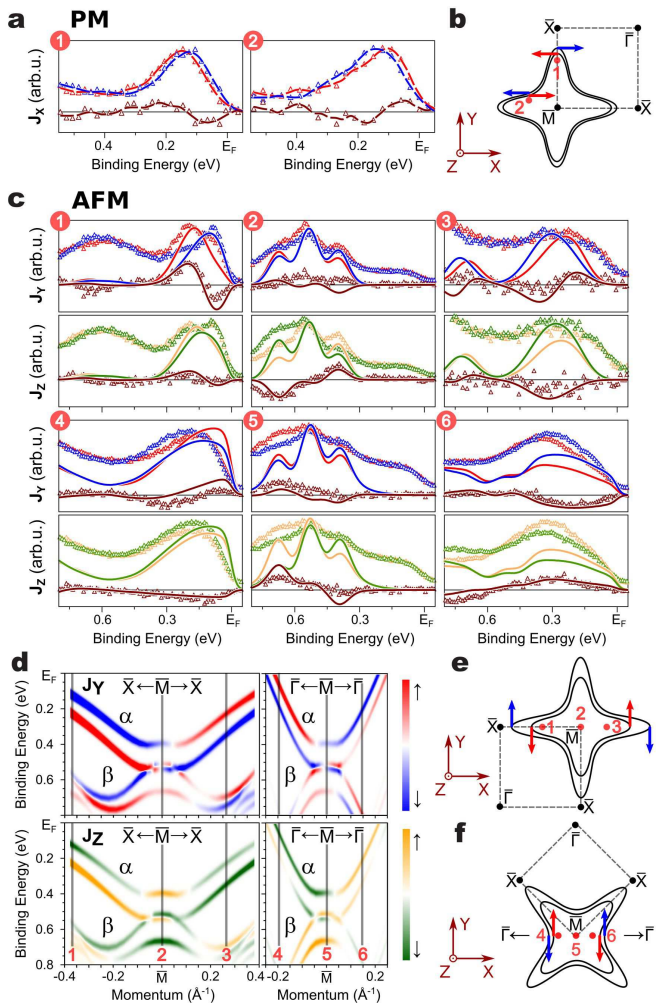}
\caption{\label{fig:2}
SR-ARPES measurements of state $\alpha$ for PM (a) and AFM (c) phases of TRS.
Measured {\it k}-points are denoted 1 and 2 for PM (b) and 1--6 for AFM (e)--(f).
Red and blue symbols show spin-up $J^{\uparrow}$ and down $J^{\downarrow}$ EDCs
for $J_X$~(a) and for $J_Y$~(c). Yellow and green symbols in (c) are $J_Z^{\uparrow}$ and $J_Z^{\downarrow}$.
Brown symbols are the net spin $J^{\uparrow}-J^{\downarrow}$. Lines in graph (a) are guide for the eye. Calculated EDCs for AFM (c) are shown with lines of the same color code. EDCs are energy broadened and $k$-averaged along the analyzer slit to mimic the experimental energy resolution and acceptance angle. Energy-momentum maps in graph (d) show calculated net spin-photocurrent without broadening and averaging. Theoretical data in (c) and (d) are averaged over the opposite magnetization directions. The $XYZ$ axes of the analyzer relative to SBZ are shown in graphs (b), (e), and (f) by brown arrows, see SM~\cite{Note1} for details of geometry. In (b), (e), and (f), arrows (red for inner and blue for outer CEC) are a guide to relate EDCs in (a) and (c) to spin polarization in ground state.}
\end{figure}
%%%%%%%%%%%%%%%%%%%%%%%%%%%%

The DFT band structure alongside the ARPES from the Si-terminated surface of TRS is shown for PM in Fig.~\ref{fig:1}(b) and for AFM in Fig.~\ref{fig:1}(c). The three surface states are labeled $\alpha$, $\beta$, and $\gamma$. The occupied states $\alpha$ and $\beta$ are clearly seen in the experiment in perfect agreement with the theory. For the PM phase, the states $\alpha$, $\beta$, and $\gamma$ have only in-plane spin components $S_x$ and $S_y$. The splitting obviously stems from the SOI of Rh in the non-centrosymmetric Si--Rh--Si--Tb surface block~\cite{Generalov2018, Nechaev_PRB_2018, HoRhSi}. The inset in the $\bar{M}$-$\bar{X}$ graph of Fig.~\ref{fig:1}(b) shows a magnified ARPES map of  $\alpha$  with a well-resolved ${\mathbf k}$-dependent splitting, which reaches 35~meV at the Fermi level (see the black arrows). In Figs.~\ref{fig:1}(b) and~\ref{fig:1}(c), the color indicates the orientation of the in-plane spin: clockwise (red) or anti-clockwise (blue) assuming the origin at $\bar{M}$, see the sketch of the SBZ in Fig.~\ref{fig:1}(a). Thus, the line $\bar{M}$-$\bar{\Gamma}$ is related to $\bar{M}$-$\bar{X}$ by an anti-clockwise rotation around $\bar{M}$ by $\pi/4$, whereby  $\beta$ and $\gamma$ preserve their chirality (a classical Rashba behavior), whereas the chirality of $\alpha$ reverses: along $\bar{M}$-$\bar{\Gamma}$ the inner branch becomes red and the outer becomes blue, Fig.~\ref{fig:1}(c), indicating the rotation of spin by $3\pi/4$.

The comparison of Figs.~\ref{fig:1}(b) and \ref{fig:1}(c) suggests that the in-plane spin polarization of the states $\alpha$, $\beta$, and $\gamma$ survives the transition from PM to AFM phase, i.e., the emergence of the FM order within the Si--Rh--Si--Tb surface block. The experimentally observed large Zeeman-like splitting [Fig.~\ref{fig:1}(c)] and the perfect symmetry of the split contours [Fig.~\ref{fig:1}(a)] points to the out-of-plane orientation of the Tb 4$f$ moments~\cite{HoRhSi}. This gives rise to a sizable out-of-plane spin component $S_z$ of  $\alpha$ and $\beta$  and strongly enhances their splitting, cf. Figs.~\ref{fig:1}(b) and \ref{fig:1}(c), i.e., the exchange field felt by these states is much stronger than the SOI. The splitting of  $\gamma$  is only slightly affected by the magnetization because of the negligible overlap with Tb: the state $\gamma$ couples to Tb 4$f$s only via a weak indirect exchange~\cite{Nechaev_PRB_2018}.

We now consider the unusual spin structure of the state $\alpha$. The calculated spin orientation over the CECs around $\bar{M}$ for both PM and AFM phases is shown in Figs.~\ref{fig:1}(d) and~\ref{fig:1}(e). For the PM phase, the spin lies in plane with a curious triple winding around the fourfold-symmetric contours. In Fig.~\ref{fig:1}(d), the points between which the spin rotates by $2\pi$ are shown by the encircled arrows. For the AFM phase, the sizable $S_z$  is due to the large out-of-plane magnetization , Fig.~\ref{fig:1}(e). At the same time, in spite of a relatively weak SOI on Rh atoms the unique spin-momentum locking survives and remains practically unaltered when the magnetic order sets in.

The essence of the triple winding is a fast rotation of the in-plane spin $\mathrm{\mathbf{S}}_{\parallel}= S_x\hat{\mathrm{\mathbf{x}}} + S_y\hat{\mathrm{\mathbf{y}}}$ over a symmetry-irreducible fragment of CEC between the directions $\bar{M}$-$\bar{X}$ and $\bar{M}$-$\bar{\Gamma}$. Experimentally, we are limited by the fact that the spin-quantization axes $X$, $Y$, and $Z$ of the spin analyzer depend on the emission direction $\mathbf{J}$: $X$ and $Y$ are perpendicular and $Z$ is parallel to $\mathbf{J}$, the respective components of the photocurrent being $J_X$, $J_Y$, and $J_Z$. Thus, it is important to select the ${\mathbf k}$-points such that the actual  direction of $\mathrm{\mathbf{S}}_{\parallel}$ be close to $X$ or $Y$ axis and, at the same time, unambiguously manifest the effect.  In particular, in Fig.~\ref{fig:2}(a), the energy distribution curves (EDC) for the PM phase at two characteristic ${\mathbf k}$-points show the flip of the $J_X$-polarization, i.e., of the net-spin $X$ photocurrent.

In the AFM phase, the magnetization  gives rise to a large  $S_z$, which for each branch of $\alpha$ or $\beta$ does not change sign along $\bar{X}$-$\bar{M}$-$\bar{X}$ and $\bar{\Gamma}$-$\bar{M}$-$\bar{\Gamma}$. This is expected to manifest itself in a sizable $J_Z$-polarization of the same sign. In the SR-ARPES measurements, depending on the tilt and polar angles, the large $S_z$ also contributes to $J_X$ and $J_Y$. In turn, $J_Z$ may be influenced by $\mathrm{\mathbf{S}}_{\parallel}$. An additional contribution to the $J_Z$ polarization may be caused by the light breaking the symmetry of the experiment. This is a matrix element effect (MEE), which depends on the light incidence and on the final state~\cite{Bentmann2017} , and the present photoemission theory is instrumental in order to separate the different sources of $J_Z$.

However, the observed behavior of $J_Z$ cannot be reconciled with the assumption that the signal comes from a single magnetic domain: the $J_Z$-polarization is weak, and at $\bar{M}$ it changes sign depending on the experimental geometry, cf. points 2 [$\bar{X}$-$\bar{M}$-$\bar{X}$ setup, Fig.~\ref{fig:2}(e)] and 5 [$\bar{\Gamma}$-$\bar{M}$-$\bar{\Gamma}$ setup, Fig.~\ref{fig:2}(f)] in Fig.~\ref{fig:2}(c). Furthermore, $J_Z$ has opposite polarization in points 1~(4) and 3~(6). According to our theory, Fig.~\ref{fig:2}(d), this implies a presence of oppositely magnetized domains, i.e., in the ground state the averaging over magnetization destroys the out-of-plane spin, thereby eliminating the $S_z$ contribution to $J_Z$. Also the $\mathrm{\mathbf{S}}_{\parallel}$ contribution to $J_Z$ turns out to be rather weak in the AFM phase not only in the vicinity of $\bar{M}$ but also far from it. Thus, the $J_Z$ EDCs in Fig.~\ref{fig:2}(c) are a manifestation of the MEE, since the influence of the spin in the ground state is here unimportant.

Another manifestation of the MEE is that over a rather wide range around $\bar{M}$ the $J_Y$-polarization strongly deviates from the spin of the initial state, cf.~Fig.~\ref{fig:2}(d) and \ref{fig:1}(c), while sufficiently far from $\bar{M}$ it follows  $\mathrm{\mathbf{S}}_{\parallel}$ of the initial state, both in the theory and in the experiment. Thus, we unambiguously relate the measured polarization of the photocurrent with the ground-state spin structure. The good agreement between the theoretical and measured $J_Y$ and $J_Z$ in Fig.~\ref{fig:2}(c) fully supports the triple-winding interpretation of the measured spin structure. Interestingly, among the vast variety of known Rashba systems with stronger or weaker pronounced spin patterns of various shape this specific pattern has never been observed before.

\begin{figure}[t]
\centering
\includegraphics[width=\columnwidth]{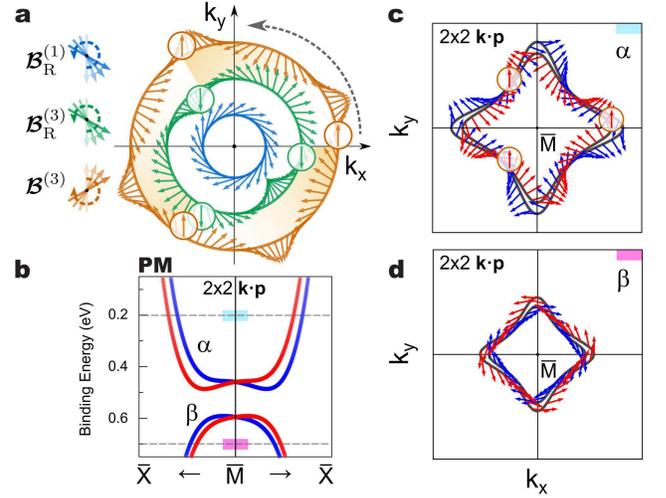}
\caption{(a)  Effective spin-orbit magnetic fields  ${\bm{\mathcal{B}}}_{\rtm{R}}^{(1)}$ (linear Rashba effect, blue arrows),  ${\bm{\mathcal{B}}}_{\rtm{R}}^{(3)}$ (``heavy-hole'' cubic Rashba effect, green arrows), and  ${\bm{\mathcal{B}}}^{(3)}$ (cubic Rashba effect for the state $\alpha$, orange arrows) as a function of polar angle $\varphi_{{\mathbf k}}$. The legends on the left show the rotation direction of the vectors ${\bm{\mathcal{B}}}_{\rtm{R}}^{(1)}$, ${\bm{\mathcal{B}}}_{\rtm{R}}^{(3)}$, and  ${\bm{\mathcal{B}}}^{(3)}$ with the anti-clockwise rotation of ${\mathbf k}$ (gray dashed arc arrow). (b) Band structure of the states $\alpha$ and $\beta$ by the Hamiltonian~(\ref{TwoBandHam}) presented by red ($S_y>0$) and blue ($S_y<0$) bands.
Graphs (c) and (d) show spin-resolved CECs by the ${\mathbf k}\cdot{\mathbf p}$ model at the binding energy of 0.2 and 0.7~eV, respectively [indicated by dashed gray lines in graph (b)], with the in-plane spin orientation shown by red (blue) arrows for the inner contour and by blue (red) for outer contour of the state $\alpha$ ($\beta$).}
\label{fig:3}
\end{figure}

To prove that the observed spin pattern is a manifestation of the cubic Rashba effect let us focus on the PM phase. In the literature, the term ``cubic'' implies a specific form of the two-band Rashba effective Hamiltonian for the total angular momentum states $|jm_j\rangle$ with the $z$ projection $m_j=\pm3/2$ (in particular for 2D heavy holes)~\cite{Gerchikov_SPS_1992, Winkler_PRB_2000, Winkler_SST_2008}. There, the cubic spin-orbit splitting of these states is due to the term $H_{\rtm{R}}^{(3)}=i\widetilde{\gamma}(k_-^3\sigma_+ - k_+^3\sigma_-)$, where $k_{\pm}=k_x\pm ik_y$, $\sigma_{\pm}=(\sigma_x\pm i\sigma_y)/2$, and $\sigma_{\rho}$ ($\rho=x$, $y$, and $z$) are Pauli matrices referring to a {\it pseudospin} (or {\it effective spin}) since they operate on the $|jm_j\rangle$ states, as emphasized in Refs.~\cite{Schliemann_PRB_2005, Bladwell_PRB_2015}. To highlight the impact of the cubic Rashba effect on the {\it pseudospin}, the ${\mathbf k}$-cubic contribution is expressed as a Zeeman-like term $H_{\rtm{R}}^{(3)}=\widetilde{\gamma}{\bm\sigma} \cdot {\bm{\mathcal{B}}}_{\rtm{R}}^{(3)}$, where ${\bm{\mathcal{B}}}_{\rtm{R}}^{(3)} =  k^3(\sin3\varphi_{{\mathbf k}}, -\cos3\varphi_{{\mathbf k}}, 0)$ is the EMF with $k=\sqrt{k_x^2+k_y^2}$ and $\varphi_{{\mathbf k}}$ being the polar angle of the momentum ${\mathbf k}$. Thus, the EMF drives {\it pseudospin} to be collinear with ${\bm{\mathcal{B}}}_{\rtm{R}}^{(3)}$ at a given ${\mathbf k}$, see Fig.~\ref{fig:3}(a)~\cite{Winkler_KP}. As a result, the triple winding of the {\it pseudospins} with a complete $2\pi$ rotation of ${\mathbf k}$ is a hallmark of the cubic effect. In contrast, the classical (linear) Rashba two-band Hamiltonian is $H_{\rtm{R}}^{(1)}=i\widetilde{\alpha}(k_-\sigma_+ - k_+\sigma_-)=\widetilde{\alpha}{\bm\sigma}\cdot {\bm{\mathcal{B}}}_{\rtm{R}}^{(1)}$ with a single winding of the EMF ${\bm{\mathcal{B}}}_{\rtm{R}}^{(1)} =  k(\sin\varphi_{{\mathbf k}},-\cos\varphi_{{\mathbf k}},0)$, Fig.~\ref{fig:3}(a).

%+++++++++++++++++++++++++++++++++++++++++++++++++++++++++++++++++++++++++++++++++
%+++++++++++++++++++++++++++++++++++++++++++++++++++++++++++++++++++++++++++++++++

In order to understand whether the observed triple winding of the true spin is a manifestation of the cubic Rashba effect,
we draw on the {\it microscopic} approach of Refs.~\cite{Nechaev_PRBR_2016, Nechaev_PRB_2018, Nechaev_PRB_2019} to derive a two-band ${\mathbf k}\cdot{\mathbf p}$ Hamiltonian of the form (see the SM~\cite{Note1})
%-------------------------------------------------------------------------------------
\begin{eqnarray}\label{TwoBandHam}
   H^{2\times2}_{\rtm{\mathbf{kp}}}&=& [\epsilon + M(k)k^2 + W(k_+^4+k_-^4)] \sigma_0 \nonumber\\
   &+&i\widetilde{\alpha}(k)(k_-\sigma_+-k_+\sigma_-)\\
   &+&i\widetilde{\gamma}(k_-^3\sigma_--k_+^3\sigma_+).\nonumber
\end{eqnarray}
%-------------------------------------------------------------------------------------
Here, $\epsilon$ is the surface-state energy at $\bar{M}$. The effective-mass term $M(k)=M^{(2)}+M^{(4)}k^2$ contains a second order correction, and the parameter $W$ describes the fourfold warping of the CECs. In Eq.~(\ref{TwoBandHam}), the renormalized Rashba parameter $\widetilde{\alpha}(k)=\widetilde{\alpha}^{(1)}+\widetilde{\alpha}^{(3)}k^2$ accounts for the conventional (orthogonal) in-plane spin-momentum locking governed by ${\bm{\mathcal{B}}}_{\rtm{R}}^{(1)}$. The term with $\widetilde{\gamma}$ brings about the cubic Rashba effect we are interested in. Note that the respective EMF $\bm{\mathcal{B}}^{(3)} = k^3(\sin3\varphi_{{\mathbf k}}, \cos3\varphi_{{\mathbf k}}, 0)$ has a fourfold symmetry, whereas the ``heavy-hole'' field ${\bm{\mathcal{B}}}_{\rtm{R}}^{(3)}$ has a twofold symmetry, see Fig.~\ref{fig:3}(a).

The spin polarization of the states $\alpha$ and $\beta$ rapidly grows in moving away from $\bar{M}$: being practically unpolarized in the vicinity of $\bar{M}$ they become almost completely polarized far from it, Fig.~\ref{fig:1}(c). This is due to the interaction between $\alpha$ and $\beta$, and, naturally, the $2\times 2$ Hamiltonian~(\ref{TwoBandHam}) does not have this property. However, if we associate the $\sigma$ matrices in Eq.~(\ref{TwoBandHam}) with the spin~\cite{Note1}, Fig.~\ref{fig:3}(b), the Hamiltonian turns out to accurately describe both the triple winding in the $\alpha$ contour, Fig.~\ref{fig:3}(c), and the single winding in the $\beta$ contour, Fig.~\ref{fig:3}(d).

%-----------------------------------------------------------------------------------------------------------------------------------
\begin{table}
\caption{\label{tab:table1}Parameters of the Hamiltonian~(\ref{TwoBandHam}) for the states $\alpha$ and $\beta$ in Rydberg atomic units: $\hbar=2m_0=e^2/2=1$.}
\begin{ruledtabular}
\begin{tabular}{lcc}
  Surface state                           & $\alpha$                     & $\beta$   \\
 \hline
$\epsilon/M^{(2)}$                                  & -0.035/-1.55                            & -0.045/-0.96  \\
$M^{(4)}/W$                                  & 3.93$\times10^3$/-1.43$\times10^3$    &  -3.93$\times10^3$/1.43$\times10^3$ \\
$\widetilde{\alpha}^{(1)}/\widetilde{\alpha}^{(3)}/\widetilde{\gamma}$      &   0.039/-11.85/-23.35           & 0.039/6.05/3.30    \\
\end{tabular}
\end{ruledtabular}
\end{table}
%-----------------------------------------------------------------------------------------------------------------------------------

The spin of the state $\alpha$ follows the winding of the field $\bm{\mathcal{B}}^{(3)}$, Fig.~\ref{fig:3}(c), revealing the dominance of the $k_{\pm}$-cubic term over the $k_{\pm}$-linear one. Actually, for this state $\widetilde{\gamma}$ is almost twice as large as $\widetilde{\alpha}^{(3)}$ (the {\it microscopically} obtained parameters are listed in Table~\ref{tab:table1}). By contrast, for the state $\beta$ the $k_{\pm}$-linear term dominates because here $\widetilde{\gamma}$ is two times smaller than $\widetilde{\alpha}^{(3)}$, Table~\ref{tab:table1}. This results in a single spin winding around the CEC, Fig.~\ref{fig:3}(d). Thus, we conclude that the observed spin structure can be thought of as a superposition of a linear and a cubic Rashba effects, which compete to produce the winding of surface-state spin shown in Fig.~\ref{fig:1}(d).

In summary, with SR-ARPES we unveiled an unusual in-plane spin structure of surface states with a triple winding of the spin along the constant energy contours in the exemplary antiferromagnet TbRh$_2$Si$_2$, a representative of a wide class of rare-earth ternary materials. The unique spin structure appears to be rather robust and persists when a strong out-of-plane magnetic order of the Tb moments sets in. This property is due to a cubic Rashba effect, whose strength can be tuned by a proper choice of the transition metal atom. Combined with a different orientation and strength of the exchange field near the surface it can create a variety of fascinating spin patterns. A fundamentally important finding is that relatively light atoms like Rh may give rise to a distinct spin-momentum gyration stable in an exchange field much stronger than the spin-orbit field.

\begin{acknowledgments}
This work was supported by the German Research Foundation (KR-3831/5-1, GRK1621, Fermi-NEst, and SFB1143), the Spanish Ministry of Science, Innovation  and  Universities (Project No. FIS2016-76617-P and MAT-2017-88374-P) and Funding from the Department of Education of the Basque Government under Grant No. IT1164-19. St. Petersburg State University (Grant No. 40990069) and the Russian Foundation for Basic Research (Grant No. 20-32-70127) are acknowledged.
\end{acknowledgments}


\begin{thebibliography}{42}%
\makeatletter
\providecommand \@ifxundefined [1]{%
 \@ifx{#1\undefined}
}%
\providecommand \@ifnum [1]{%
 \ifnum #1\expandafter \@firstoftwo
 \else \expandafter \@secondoftwo
 \fi
}%
\providecommand \@ifx [1]{%
 \ifx #1\expandafter \@firstoftwo
 \else \expandafter \@secondoftwo
 \fi
}%
\providecommand \natexlab [1]{#1}%
\providecommand \enquote  [1]{``#1''}%
\providecommand \bibnamefont  [1]{#1}%
\providecommand \bibfnamefont [1]{#1}%
\providecommand \citenamefont [1]{#1}%
\providecommand \href@noop [0]{\@secondoftwo}%
\providecommand \href [0]{\begingroup \@sanitize@url \@href}%
\providecommand \@href[1]{\@@startlink{#1}\@@href}%
\providecommand \@@href[1]{\endgroup#1\@@endlink}%
\providecommand \@sanitize@url [0]{\catcode `\\12\catcode `\$12\catcode
  `\&12\catcode `\#12\catcode `\^12\catcode `\_12\catcode `\%12\relax}%
\providecommand \@@startlink[1]{}%
\providecommand \@@endlink[0]{}%
\providecommand \url  [0]{\begingroup\@sanitize@url \@url }%
\providecommand \@url [1]{\endgroup\@href {#1}{\urlprefix }}%
\providecommand \urlprefix  [0]{URL }%
\providecommand \Eprint [0]{\href }%
\providecommand \doibase [0]{http://dx.doi.org/}%
\providecommand \selectlanguage [0]{\@gobble}%
\providecommand \bibinfo  [0]{\@secondoftwo}%
\providecommand \bibfield  [0]{\@secondoftwo}%
\providecommand \translation [1]{[#1]}%
\providecommand \BibitemOpen [0]{}%
\providecommand \bibitemStop [0]{}%
\providecommand \bibitemNoStop [0]{.\EOS\space}%
\providecommand \EOS [0]{\spacefactor3000\relax}%
\providecommand \BibitemShut  [1]{\csname bibitem#1\endcsname}%
\let\auto@bib@innerbib\@empty
%</preamble>
\bibitem [{\citenamefont {Schulz}\ \emph {et~al.}(2019)\citenamefont {Schulz},
  \citenamefont {Nechaev}, \citenamefont {G\"{u}ttler}, \citenamefont
  {Poelchen}, \citenamefont {Generalov}, \citenamefont {Danzenb\"{a}cher},
  \citenamefont {Chikina}, \citenamefont {Seiro}, \citenamefont {Kliemt},
  \citenamefont {Vyazovskaya}, \citenamefont {Kim}, \citenamefont {Dudin},
  \citenamefont {Chulkov}, \citenamefont {Laubschat}, \citenamefont
  {Krasovskii}, \citenamefont {Geibel}, \citenamefont {Krellner}, \citenamefont
  {Kummer},\ and\ \citenamefont {Vyalikh}}]{Susanne2019}%
  \BibitemOpen
  \bibfield  {author} {\bibinfo {author} {\bibfnamefont {S.}~\bibnamefont
  {Schulz}}, \bibinfo {author} {\bibfnamefont {I.~A.}\ \bibnamefont {Nechaev}},
  \bibinfo {author} {\bibfnamefont {M.}~\bibnamefont {G\"{u}ttler}}, \bibinfo
  {author} {\bibfnamefont {G.}~\bibnamefont {Poelchen}}, \bibinfo {author}
  {\bibfnamefont {A.}~\bibnamefont {Generalov}}, \bibinfo {author}
  {\bibfnamefont {S.}~\bibnamefont {Danzenb\"{a}cher}}, \bibinfo {author}
  {\bibfnamefont {A.}~\bibnamefont {Chikina}}, \bibinfo {author} {\bibfnamefont
  {S.}~\bibnamefont {Seiro}}, \bibinfo {author} {\bibfnamefont
  {K.}~\bibnamefont {Kliemt}}, \bibinfo {author} {\bibfnamefont {A.~Yu.}\
  \bibnamefont {Vyazovskaya}}, \bibinfo {author} {\bibfnamefont {T.~K.}\
  \bibnamefont {Kim}}, \bibinfo {author} {\bibfnamefont {P.}~\bibnamefont
  {Dudin}}, \bibinfo {author} {\bibfnamefont {E.~V.}\ \bibnamefont {Chulkov}},
  \bibinfo {author} {\bibfnamefont {C.}~\bibnamefont {Laubschat}}, \bibinfo
  {author} {\bibfnamefont {E.~E.}\ \bibnamefont {Krasovskii}}, \bibinfo
  {author} {\bibfnamefont {C.}~\bibnamefont {Geibel}}, \bibinfo {author}
  {\bibfnamefont {C.}~\bibnamefont {Krellner}}, \bibinfo {author}
  {\bibfnamefont {K.}~\bibnamefont {Kummer}}, \ and\ \bibinfo {author}
  {\bibfnamefont {D.~V.}\ \bibnamefont {Vyalikh}},\ }\bibfield  {title}
  {\enquote {\bibinfo {title} {{Emerging 2D-ferromagnetism and strong
  spin-orbit coupling at the surface of valence-fluctuating EuIr$_2$Si$_2$}},}\
  }\href {\doibase 10.1038/s41535-019-0166-z} {\bibfield  {journal} {\bibinfo
  {journal} {npj Quantum Mater.}\ }\textbf {\bibinfo {volume} {4}},\ \bibinfo
  {pages} {26} (\bibinfo {year} {2019})}\BibitemShut {NoStop}%
\bibitem [{\citenamefont {Generalov}\ \emph {et~al.}(2018)\citenamefont
  {Generalov}, \citenamefont {Falke}, \citenamefont {Nechaev}, \citenamefont
  {Otrokov}, \citenamefont {G\"{u}ttler}, \citenamefont {Chikina},
  \citenamefont {Kliemt}, \citenamefont {Seiro}, \citenamefont {Kummer},
  \citenamefont {Danzenb\"{a}cher}, \citenamefont {Usachov}, \citenamefont
  {Kim}, \citenamefont {Dudin}, \citenamefont {Chulkov}, \citenamefont
  {Laubschat}, \citenamefont {Geibel}, \citenamefont {Krellner},\ and\
  \citenamefont {Vyalikh}}]{Generalov2018}%
  \BibitemOpen
  \bibfield  {author} {\bibinfo {author} {\bibfnamefont {A.}~\bibnamefont
  {Generalov}}, \bibinfo {author} {\bibfnamefont {J.}~\bibnamefont {Falke}},
  \bibinfo {author} {\bibfnamefont {I.~A.}\ \bibnamefont {Nechaev}}, \bibinfo
  {author} {\bibfnamefont {M.~M.}\ \bibnamefont {Otrokov}}, \bibinfo {author}
  {\bibfnamefont {M.}~\bibnamefont {G\"{u}ttler}}, \bibinfo {author}
  {\bibfnamefont {A.}~\bibnamefont {Chikina}}, \bibinfo {author} {\bibfnamefont
  {K.}~\bibnamefont {Kliemt}}, \bibinfo {author} {\bibfnamefont
  {S.}~\bibnamefont {Seiro}}, \bibinfo {author} {\bibfnamefont
  {K.}~\bibnamefont {Kummer}}, \bibinfo {author} {\bibfnamefont
  {S.}~\bibnamefont {Danzenb\"{a}cher}}, \bibinfo {author} {\bibfnamefont
  {D.}~\bibnamefont {Usachov}}, \bibinfo {author} {\bibfnamefont {T.~K.}\
  \bibnamefont {Kim}}, \bibinfo {author} {\bibfnamefont {P.}~\bibnamefont
  {Dudin}}, \bibinfo {author} {\bibfnamefont {E.~V.}\ \bibnamefont {Chulkov}},
  \bibinfo {author} {\bibfnamefont {C.}~\bibnamefont {Laubschat}}, \bibinfo
  {author} {\bibfnamefont {C.}~\bibnamefont {Geibel}}, \bibinfo {author}
  {\bibfnamefont {C.}~\bibnamefont {Krellner}}, \ and\ \bibinfo {author}
  {\bibfnamefont {D.~V.}\ \bibnamefont {Vyalikh}},\ }\bibfield  {title}
  {\enquote {\bibinfo {title} {{Strong spin-orbit coupling in the
  noncentrosymmetric Kondo lattice}},}\ }\href {\doibase
  10.1103/PhysRevB.98.115157} {\bibfield  {journal} {\bibinfo  {journal} {Phys.
  Rev. B}\ }\textbf {\bibinfo {volume} {98}},\ \bibinfo {pages} {115157}
  (\bibinfo {year} {2018})}\BibitemShut {NoStop}%
\bibitem [{\citenamefont {Nechaev}\ and\ \citenamefont
  {Krasovskii}(2018)}]{Nechaev_PRB_2018}%
  \BibitemOpen
  \bibfield  {author} {\bibinfo {author} {\bibfnamefont {I.~A.}\ \bibnamefont
  {Nechaev}}\ and\ \bibinfo {author} {\bibfnamefont {E.~E.}\ \bibnamefont
  {Krasovskii}},\ }\bibfield  {title} {\enquote {\bibinfo {title}
  {{Relativistic splitting of surface states at Si-terminated surfaces of the
  layered intermetallic compounds $R{T}_{2}{\mathrm{Si}}_{2}$ ($R$=rare earth;
  $T$=Ir, Rh)}},}\ }\href {\doibase 10.1103/PhysRevB.98.245415} {\bibfield
  {journal} {\bibinfo  {journal} {Phys. Rev. B}\ }\textbf {\bibinfo {volume}
  {98}},\ \bibinfo {pages} {245415} (\bibinfo {year} {2018})}\BibitemShut
  {NoStop}%
\bibitem [{\citenamefont {Nechaev}\ and\ \citenamefont
  {Krasovskii}(2019)}]{Nechaev_PRB_2019}%
  \BibitemOpen
  \bibfield  {author} {\bibinfo {author} {\bibfnamefont {I.~A.}\ \bibnamefont
  {Nechaev}}\ and\ \bibinfo {author} {\bibfnamefont {E.~E.}\ \bibnamefont
  {Krasovskii}},\ }\bibfield  {title} {\enquote {\bibinfo {title} {{Spin
  polarization by first-principles relativistic $\mathrm{k}\cdot\mathrm{p}$
  theory: Application to the surface alloys ${\mathrm{PbAg}}_{2}$ and
  ${\mathrm{BiAg}}_{2}$}},}\ }\href {\doibase 10.1103/PhysRevB.100.115432}
  {\bibfield  {journal} {\bibinfo  {journal} {Phys. Rev. B}\ }\textbf {\bibinfo
  {volume} {100}},\ \bibinfo {pages} {115432} (\bibinfo {year}
  {2019})}\BibitemShut {NoStop}%
\bibitem [{\citenamefont {Bondarenko}\ \emph {et~al.}(2013)\citenamefont
  {Bondarenko}, \citenamefont {Gruznev}, \citenamefont {Yakovlev},
  \citenamefont {Tupchaya}, \citenamefont {Usachov}, \citenamefont {Vilkov},
  \citenamefont {Fedorov}, \citenamefont {Vyalikh}, \citenamefont {Eremeev},
  \citenamefont {Chulkov}, \citenamefont {Zotov},\ and\ \citenamefont
  {Saranin}}]{Bondarenko_SciRep_2013}%
  \BibitemOpen
  \bibfield  {author} {\bibinfo {author} {\bibfnamefont {L.~V.}\ \bibnamefont
  {Bondarenko}}, \bibinfo {author} {\bibfnamefont {D.~V.}\ \bibnamefont
  {Gruznev}}, \bibinfo {author} {\bibfnamefont {A.~A.}\ \bibnamefont
  {Yakovlev}}, \bibinfo {author} {\bibfnamefont {A.~Y.}\ \bibnamefont
  {Tupchaya}}, \bibinfo {author} {\bibfnamefont {D.}~\bibnamefont {Usachov}},
  \bibinfo {author} {\bibfnamefont {O.}~\bibnamefont {Vilkov}}, \bibinfo
  {author} {\bibfnamefont {A.}~\bibnamefont {Fedorov}}, \bibinfo {author}
  {\bibfnamefont {D.~V.}\ \bibnamefont {Vyalikh}}, \bibinfo {author}
  {\bibfnamefont {S.~V.}\ \bibnamefont {Eremeev}}, \bibinfo {author}
  {\bibfnamefont {E.~V.}\ \bibnamefont {Chulkov}}, \bibinfo {author}
  {\bibfnamefont {A.~V.}\ \bibnamefont {Zotov}}, \ and\ \bibinfo {author}
  {\bibfnamefont {A.~A.}\ \bibnamefont {Saranin}},\ }\bibfield  {title}
  {\enquote {\bibinfo {title} {Large spin splitting of metallic surface-state
  bands at adsorbate-modified gold/silicon surfaces},}\ }\href {\doibase
  10.1038/srep01826} {\bibfield  {journal} {\bibinfo  {journal} {Sci. Rep.}\
  }\textbf {\bibinfo {volume} {3}},\ \bibinfo {pages} {1826} (\bibinfo {year}
  {2013})}\BibitemShut {NoStop}%
\bibitem [{\citenamefont {Friedrich}\ \emph {et~al.}(2017)\citenamefont
  {Friedrich}, \citenamefont {Caciuc}, \citenamefont {Bihlmayer}, \citenamefont
  {Atodiresei},\ and\ \citenamefont {Bl\"ugel}}]{Friedrich_NJP_2017}%
  \BibitemOpen
  \bibfield  {author} {\bibinfo {author} {\bibfnamefont {Rico}\ \bibnamefont
  {Friedrich}}, \bibinfo {author} {\bibfnamefont {Vasile}\ \bibnamefont
  {Caciuc}}, \bibinfo {author} {\bibfnamefont {Gustav}\ \bibnamefont
  {Bihlmayer}}, \bibinfo {author} {\bibfnamefont {Nicolae}\ \bibnamefont
  {Atodiresei}}, \ and\ \bibinfo {author} {\bibfnamefont {Stefan}\ \bibnamefont
  {Bl\"ugel}},\ }\bibfield  {title} {\enquote {\bibinfo {title} {{Designing the
  Rashba spin texture by adsorption of inorganic molecules}},}\ }\href
  {\doibase 10.1088/1367-2630/aa64a1} {\bibfield  {journal} {\bibinfo
  {journal} {New J. Phys.}\ }\textbf {\bibinfo {volume} {19}},\ \bibinfo
  {pages} {043017} (\bibinfo {year} {2017})}\BibitemShut {NoStop}%
\bibitem [{\citenamefont {Moriya}\ \emph {et~al.}(2014)\citenamefont {Moriya},
  \citenamefont {Sawano}, \citenamefont {Hoshi}, \citenamefont {Masubuchi},
  \citenamefont {Shiraki}, \citenamefont {Wild}, \citenamefont {Neumann},
  \citenamefont {Abstreiter}, \citenamefont {Bougeard}, \citenamefont {Koga},\
  and\ \citenamefont {Machida}}]{MoriyaPRL_2014}%
  \BibitemOpen
  \bibfield  {author} {\bibinfo {author} {\bibfnamefont {Rai}\ \bibnamefont
  {Moriya}}, \bibinfo {author} {\bibfnamefont {Kentarou}\ \bibnamefont
  {Sawano}}, \bibinfo {author} {\bibfnamefont {Yusuke}\ \bibnamefont {Hoshi}},
  \bibinfo {author} {\bibfnamefont {Satoru}\ \bibnamefont {Masubuchi}},
  \bibinfo {author} {\bibfnamefont {Yasuhiro}\ \bibnamefont {Shiraki}},
  \bibinfo {author} {\bibfnamefont {Andreas}\ \bibnamefont {Wild}}, \bibinfo
  {author} {\bibfnamefont {Christian}\ \bibnamefont {Neumann}}, \bibinfo
  {author} {\bibfnamefont {Gerhard}\ \bibnamefont {Abstreiter}}, \bibinfo
  {author} {\bibfnamefont {Dominique}\ \bibnamefont {Bougeard}}, \bibinfo
  {author} {\bibfnamefont {Takaaki}\ \bibnamefont {Koga}}, \ and\ \bibinfo
  {author} {\bibfnamefont {Tomoki}\ \bibnamefont {Machida}},\ }\bibfield
  {title} {\enquote {\bibinfo {title} {{Cubic Rashba Spin-Orbit Interaction of
  a Two-Dimensional Hole Gas in a Strained-$\mathrm{Ge}/\mathrm{SiGe}$ Quantum
  Well}},}\ }\href {\doibase 10.1103/PhysRevLett.113.086601} {\bibfield
  {journal} {\bibinfo  {journal} {Phys. Rev. Lett.}\ }\textbf {\bibinfo
  {volume} {113}},\ \bibinfo {pages} {086601} (\bibinfo {year}
  {2014})}\BibitemShut {NoStop}%
\bibitem [{\citenamefont {Liu}\ \emph {et~al.}(2018)\citenamefont {Liu},
  \citenamefont {Marcellina}, \citenamefont {Hamilton},\ and\ \citenamefont
  {Culcer}}]{Hong_PRL_2018}%
  \BibitemOpen
  \bibfield  {author} {\bibinfo {author} {\bibfnamefont {Hong}\ \bibnamefont
  {Liu}}, \bibinfo {author} {\bibfnamefont {E.}~\bibnamefont {Marcellina}},
  \bibinfo {author} {\bibfnamefont {A.~R.}\ \bibnamefont {Hamilton}}, \ and\
  \bibinfo {author} {\bibfnamefont {Dimitrie}\ \bibnamefont {Culcer}},\
  }\bibfield  {title} {\enquote {\bibinfo {title} {{Strong Spin-Orbit
  Contribution to the Hall Coefficient of Two-Dimensional Hole Systems}},}\
  }\href {\doibase 10.1103/PhysRevLett.121.087701} {\bibfield  {journal}
  {\bibinfo  {journal} {Phys. Rev. Lett.}\ }\textbf {\bibinfo {volume} {121}},\
  \bibinfo {pages} {087701} (\bibinfo {year} {2018})}\BibitemShut {NoStop}%
\bibitem [{\citenamefont {Nakamura}\ \emph {et~al.}(2012)\citenamefont
  {Nakamura}, \citenamefont {Koga},\ and\ \citenamefont
  {Kimura}}]{Nakamura_PRL_2012}%
  \BibitemOpen
  \bibfield  {author} {\bibinfo {author} {\bibfnamefont {H.}~\bibnamefont
  {Nakamura}}, \bibinfo {author} {\bibfnamefont {T.}~\bibnamefont {Koga}}, \
  and\ \bibinfo {author} {\bibfnamefont {T.}~\bibnamefont {Kimura}},\
  }\bibfield  {title} {\enquote {\bibinfo {title} {{Experimental Evidence of
  Cubic Rashba Effect in an Inversion-Symmetric Oxide}},}\ }\href {\doibase
  10.1103/PhysRevLett.108.206601} {\bibfield  {journal} {\bibinfo  {journal}
  {Phys. Rev. Lett.}\ }\textbf {\bibinfo {volume} {108}},\ \bibinfo {pages}
  {206601} (\bibinfo {year} {2012})}\BibitemShut {NoStop}%
\bibitem [{\citenamefont {King}\ \emph {et~al.}(2014)\citenamefont {King},
  \citenamefont {McKeown~Walker}, \citenamefont {Tamai}, \citenamefont {de~la
  Torre}, \citenamefont {Eknapakul}, \citenamefont {Buaphet}, \citenamefont
  {Mo}, \citenamefont {Meevasana}, \citenamefont {Bahramy},\ and\ \citenamefont
  {Baumberger}}]{King_NatComm_2014}%
  \BibitemOpen
  \bibfield  {author} {\bibinfo {author} {\bibfnamefont {P.~D.~C.}\
  \bibnamefont {King}}, \bibinfo {author} {\bibfnamefont {S.}~\bibnamefont
  {McKeown~Walker}}, \bibinfo {author} {\bibfnamefont {A.}~\bibnamefont
  {Tamai}}, \bibinfo {author} {\bibfnamefont {A.}~\bibnamefont {de~la Torre}},
  \bibinfo {author} {\bibfnamefont {T.}~\bibnamefont {Eknapakul}}, \bibinfo
  {author} {\bibfnamefont {P.}~\bibnamefont {Buaphet}}, \bibinfo {author}
  {\bibfnamefont {S.-K.}\ \bibnamefont {Mo}}, \bibinfo {author} {\bibfnamefont
  {W.}~\bibnamefont {Meevasana}}, \bibinfo {author} {\bibfnamefont {M.~S.}\
  \bibnamefont {Bahramy}}, \ and\ \bibinfo {author} {\bibfnamefont
  {F.}~\bibnamefont {Baumberger}},\ }\bibfield  {title} {\enquote {\bibinfo
  {title} {{Quasiparticle dynamics and spin-orbital texture of the SrTiO$_3$
  two-dimensional electron gas}},}\ }\href {\doibase 10.1038/ncomms4414}
  {\bibfield  {journal} {\bibinfo  {journal} {Nature Commun.}\ }\textbf
  {\bibinfo {volume} {5}},\ \bibinfo {pages} {3414} (\bibinfo {year}
  {2014})}\BibitemShut {NoStop}%
\bibitem [{\citenamefont {Varignon}\ \emph {et~al.}(2018)\citenamefont
  {Varignon}, \citenamefont {Vila}, \citenamefont {Barth\'{e}l\'{e}my},\ and\
  \citenamefont {Bibes}}]{Varignon_NatPhys_2018}%
  \BibitemOpen
  \bibfield  {author} {\bibinfo {author} {\bibfnamefont {J.}~\bibnamefont
  {Varignon}}, \bibinfo {author} {\bibfnamefont {L.}~\bibnamefont {Vila}},
  \bibinfo {author} {\bibfnamefont {A.}~\bibnamefont {Barth\'{e}l\'{e}my}}, \
  and\ \bibinfo {author} {\bibfnamefont {M.}~\bibnamefont {Bibes}},\ }\bibfield
   {title} {\enquote {\bibinfo {title} {A new spin for oxide interfaces},}\
  }\href {\doibase 10.1038/s41567-018-0112-1} {\bibfield  {journal} {\bibinfo
  {journal} {Nat. Phys.}\ }\textbf {\bibinfo {volume} {14}},\ \bibinfo {pages}
  {322--325} (\bibinfo {year} {2018})}\BibitemShut {NoStop}%
\bibitem [{\citenamefont {Gariglio}\ \emph {et~al.}(2018)\citenamefont
  {Gariglio}, \citenamefont {Caviglia}, \citenamefont {Triscone},\ and\
  \citenamefont {Gabay}}]{Gariglio_RPP_2018}%
  \BibitemOpen
  \bibfield  {author} {\bibinfo {author} {\bibfnamefont {S.}~\bibnamefont
  {Gariglio}}, \bibinfo {author} {\bibfnamefont {A.~D.}\ \bibnamefont
  {Caviglia}}, \bibinfo {author} {\bibfnamefont {J.-M.}\ \bibnamefont
  {Triscone}}, \ and\ \bibinfo {author} {\bibfnamefont {M.}~\bibnamefont
  {Gabay}},\ }\bibfield  {title} {\enquote {\bibinfo {title} {A spin-orbit
  playground: surfaces and interfaces of transition metal oxides},}\ }\href
  {\doibase 10.1088/1361-6633/aad6ab} {\bibfield  {journal} {\bibinfo
  {journal} {Rep. Prog. Phys.}\ }\textbf {\bibinfo {volume} {82}},\ \bibinfo
  {pages} {012501} (\bibinfo {year} {2018})}\BibitemShut {NoStop}%
\bibitem [{\citenamefont {Lin}\ \emph {et~al.}(2019)\citenamefont {Lin},
  \citenamefont {Li}, \citenamefont {Do\u{g}an}, \citenamefont {Li},
  \citenamefont {Rotella}, \citenamefont {Yu}, \citenamefont {Zhang},
  \citenamefont {Li}, \citenamefont {Lew}, \citenamefont {Wang}, \citenamefont
  {Prellier}, \citenamefont {Pennycook}, \citenamefont {Chen}, \citenamefont
  {Zhong}, \citenamefont {Manchon},\ and\ \citenamefont {Wu}}]{WeinanLin2019}%
  \BibitemOpen
  \bibfield  {author} {\bibinfo {author} {\bibfnamefont {Weinan}\ \bibnamefont
  {Lin}}, \bibinfo {author} {\bibfnamefont {Lei}\ \bibnamefont {Li}}, \bibinfo
  {author} {\bibfnamefont {Fatih}\ \bibnamefont {Do\u{g}an}}, \bibinfo {author}
  {\bibfnamefont {Changjian}\ \bibnamefont {Li}}, \bibinfo {author}
  {\bibfnamefont {H\'{e}l\`{e}ne}\ \bibnamefont {Rotella}}, \bibinfo {author}
  {\bibfnamefont {Xiaojiang}\ \bibnamefont {Yu}}, \bibinfo {author}
  {\bibfnamefont {Bangmin}\ \bibnamefont {Zhang}}, \bibinfo {author}
  {\bibfnamefont {Yangyang}\ \bibnamefont {Li}}, \bibinfo {author}
  {\bibfnamefont {Wen~Siang}\ \bibnamefont {Lew}}, \bibinfo {author}
  {\bibfnamefont {Shijie}\ \bibnamefont {Wang}}, \bibinfo {author}
  {\bibfnamefont {Wilfrid}\ \bibnamefont {Prellier}}, \bibinfo {author}
  {\bibfnamefont {Stephen~J.}\ \bibnamefont {Pennycook}}, \bibinfo {author}
  {\bibfnamefont {Jingsheng}\ \bibnamefont {Chen}}, \bibinfo {author}
  {\bibfnamefont {Zhicheng}\ \bibnamefont {Zhong}}, \bibinfo {author}
  {\bibfnamefont {Aurelien}\ \bibnamefont {Manchon}}, \ and\ \bibinfo {author}
  {\bibfnamefont {Tom}\ \bibnamefont {Wu}},\ }\bibfield  {title} {\enquote
  {\bibinfo {title} {{Interface-based tuning of Rashba spin-orbit interaction
  in asymmetric oxide heterostructures with 3$d$ electrons}},}\ }\href
  {\doibase 10.1038/s41467-019-10961-z} {\bibfield  {journal} {\bibinfo
  {journal} {Nature Commun.}\ }\textbf {\bibinfo {volume} {10}},\ \bibinfo
  {pages} {3052} (\bibinfo {year} {2019})}\BibitemShut {NoStop}%
\bibitem [{\citenamefont {Gerchikov}\ and\ \citenamefont
  {Subashiev}(1992)}]{Gerchikov_SPS_1992}%
  \BibitemOpen
  \bibfield  {author} {\bibinfo {author} {\bibfnamefont {L.~G.}\ \bibnamefont
  {Gerchikov}}\ and\ \bibinfo {author} {\bibfnamefont {A.V.}\ \bibnamefont
  {Subashiev}},\ }\bibfield  {title} {\enquote {\bibinfo {title} {Spin
  splitting of size-quantization subbands in asymmetric heterostructures},}\
  }\href@noop {} {\bibfield  {journal} {\bibinfo  {journal} {Soviet physics.
  Semiconductors}\ }\textbf {\bibinfo {volume} {26}},\ \bibinfo {pages} {73}
  (\bibinfo {year} {1992})}\BibitemShut {NoStop}%
\bibitem [{\citenamefont {Winkler}(2003)}]{Winkler_KP}%
  \BibitemOpen
  \bibfield  {author} {\bibinfo {author} {\bibfnamefont {R.}~\bibnamefont
  {Winkler}},\ }\href@noop {} {\emph {\bibinfo {title} {{Spin-Orbit Coupling
  Effects in Two-Dimensional Electron and Hole Systems}}}}\ (\bibinfo
  {publisher} {Springer},\ \bibinfo {address} {Berlin},\ \bibinfo {year}
  {2003})\BibitemShut {NoStop}%
\bibitem [{\citenamefont {Marcellina}\ \emph {et~al.}(2017)\citenamefont
  {Marcellina}, \citenamefont {Hamilton}, \citenamefont {Winkler},\ and\
  \citenamefont {Culcer}}]{Marcellina_PRB_2017}%
  \BibitemOpen
  \bibfield  {author} {\bibinfo {author} {\bibfnamefont {E.}~\bibnamefont
  {Marcellina}}, \bibinfo {author} {\bibfnamefont {A.~R.}\ \bibnamefont
  {Hamilton}}, \bibinfo {author} {\bibfnamefont {R.}~\bibnamefont {Winkler}}, \
  and\ \bibinfo {author} {\bibfnamefont {Dimitrie}\ \bibnamefont {Culcer}},\
  }\bibfield  {title} {\enquote {\bibinfo {title} {{Spin-orbit interactions in
  inversion-asymmetric two-dimensional hole systems: A variational
  analysis}},}\ }\href {\doibase 10.1103/PhysRevB.95.075305} {\bibfield
  {journal} {\bibinfo  {journal} {Phys. Rev. B}\ }\textbf {\bibinfo {volume}
  {95}},\ \bibinfo {pages} {075305} (\bibinfo {year} {2017})}\BibitemShut
  {NoStop}%
\bibitem [{\citenamefont {Wang}\ and\ \citenamefont
  {Wu}(2012)}]{Wang_PRB_2012}%
  \BibitemOpen
  \bibfield  {author} {\bibinfo {author} {\bibfnamefont {L.}~\bibnamefont
  {Wang}}\ and\ \bibinfo {author} {\bibfnamefont {M.~W.}\ \bibnamefont {Wu}},\
  }\bibfield  {title} {\enquote {\bibinfo {title} {{Hole spin relaxation in
  $p$-type (111) GaAs quantum wells}},}\ }\href {\doibase
  10.1103/PhysRevB.85.235308} {\bibfield  {journal} {\bibinfo  {journal} {Phys.
  Rev. B}\ }\textbf {\bibinfo {volume} {85}},\ \bibinfo {pages} {235308}
  (\bibinfo {year} {2012})}\BibitemShut {NoStop}%
\bibitem [{\citenamefont {Kondo}(2015)}]{Kondo_NJP_2015}%
  \BibitemOpen
  \bibfield  {author} {\bibinfo {author} {\bibfnamefont {Kenji}\ \bibnamefont
  {Kondo}},\ }\bibfield  {title} {\enquote {\bibinfo {title} {{Spin filter
  effects in an Aharonov-Bohm ring with double quantum dots under general
  Rashba spin-orbit interactions}},}\ }\href {\doibase
  10.1088/1367-2630/18/1/013002} {\bibfield  {journal} {\bibinfo  {journal}
  {New J. Phys.}\ }\textbf {\bibinfo {volume} {18}},\ \bibinfo {pages} {013002}
  (\bibinfo {year} {2015})}\BibitemShut {NoStop}%
\bibitem [{\citenamefont {Bladwell}\ and\ \citenamefont
  {Sushkov}(2015)}]{Bladwell_PRB_2015}%
  \BibitemOpen
  \bibfield  {author} {\bibinfo {author} {\bibfnamefont {Samuel}\ \bibnamefont
  {Bladwell}}\ and\ \bibinfo {author} {\bibfnamefont {Oleg~P.}\ \bibnamefont
  {Sushkov}},\ }\bibfield  {title} {\enquote {\bibinfo {title} {Magnetic
  focusing of electrons and holes in the presence of spin-orbit
  interactions},}\ }\href {\doibase 10.1103/PhysRevB.92.235416} {\bibfield
  {journal} {\bibinfo  {journal} {Phys. Rev. B}\ }\textbf {\bibinfo {volume}
  {92}},\ \bibinfo {pages} {235416} (\bibinfo {year} {2015})}\BibitemShut
  {NoStop}%
\bibitem [{\citenamefont {Shanavas}(2016)}]{Shanavas_PRB_2016}%
  \BibitemOpen
  \bibfield  {author} {\bibinfo {author} {\bibfnamefont {K.~V.}\ \bibnamefont
  {Shanavas}},\ }\bibfield  {title} {\enquote {\bibinfo {title} {{Theoretical
  study of the cubic Rashba effect at the ${\mathrm{SrTiO}}_{3}$(001)
  surfaces}},}\ }\href {\doibase 10.1103/PhysRevB.93.045108} {\bibfield
  {journal} {\bibinfo  {journal} {Phys. Rev. B}\ }\textbf {\bibinfo {volume}
  {93}},\ \bibinfo {pages} {045108} (\bibinfo {year} {2016})}\BibitemShut
  {NoStop}%
\bibitem [{\citenamefont {Felner}\ and\ \citenamefont
  {Nowik}(1984)}]{Felner_JPCS_1984}%
  \BibitemOpen
  \bibfield  {author} {\bibinfo {author} {\bibfnamefont {Israel}\ \bibnamefont
  {Felner}}\ and\ \bibinfo {author} {\bibfnamefont {Israel}\ \bibnamefont
  {Nowik}},\ }\bibfield  {title} {\enquote {\bibinfo {title} {Itinerant and
  local magnetism, superconductivity and mixed valency phenomena in
  {$RM_2$Si$_2$, ($R$ = rare earth, $M$ = Rh, Ru)}},}\ }\href {\doibase
  10.1016/0022-3697(84)90149-5} {\bibfield  {journal} {\bibinfo  {journal} {J.
  Phys. Chem. Solids}\ }\textbf {\bibinfo {volume} {45}},\ \bibinfo {pages}
  {419--426} (\bibinfo {year} {1984})}\BibitemShut {NoStop}%
\bibitem [{\citenamefont {Kimura}\ \emph {et~al.}(2010)\citenamefont {Kimura},
  \citenamefont {Krasovskii}, \citenamefont {Nishimura}, \citenamefont
  {Miyamoto}, \citenamefont {Kadono}, \citenamefont {Kanomaru}, \citenamefont
  {Chulkov}, \citenamefont {Bihlmayer}, \citenamefont {Shimada}, \citenamefont
  {Namatame},\ and\ \citenamefont {Taniguchi}}]{Kimura2010}%
  \BibitemOpen
  \bibfield  {author} {\bibinfo {author} {\bibfnamefont {A.}~\bibnamefont
  {Kimura}}, \bibinfo {author} {\bibfnamefont {E.~E.}\ \bibnamefont
  {Krasovskii}}, \bibinfo {author} {\bibfnamefont {R.}~\bibnamefont
  {Nishimura}}, \bibinfo {author} {\bibfnamefont {K.}~\bibnamefont {Miyamoto}},
  \bibinfo {author} {\bibfnamefont {T.}~\bibnamefont {Kadono}}, \bibinfo
  {author} {\bibfnamefont {K.}~\bibnamefont {Kanomaru}}, \bibinfo {author}
  {\bibfnamefont {E.~V.}\ \bibnamefont {Chulkov}}, \bibinfo {author}
  {\bibfnamefont {G.}~\bibnamefont {Bihlmayer}}, \bibinfo {author}
  {\bibfnamefont {K.}~\bibnamefont {Shimada}}, \bibinfo {author} {\bibfnamefont
  {H.}~\bibnamefont {Namatame}}, \ and\ \bibinfo {author} {\bibfnamefont
  {M.}~\bibnamefont {Taniguchi}},\ }\bibfield  {title} {\enquote {\bibinfo
  {title} {{Strong Rashba-Type Spin Polarization of the Photocurrent from Bulk
  Continuum States: Experiment and Theory for Bi(111)}},}\ }\href {\doibase
  10.1103/PhysRevLett.105.076804} {\bibfield  {journal} {\bibinfo  {journal}
  {Phys. Rev. Lett.}\ }\textbf {\bibinfo {volume} {105}},\ \bibinfo {pages}
  {076804} (\bibinfo {year} {2010})}\BibitemShut {NoStop}%
\bibitem [{\citenamefont {Bentmann}\ \emph {et~al.}(2017)\citenamefont
  {Bentmann}, \citenamefont {Maa\ss{}}, \citenamefont {Krasovskii},
  \citenamefont {Peixoto}, \citenamefont {Seibel}, \citenamefont {Leandersson},
  \citenamefont {Balasubramanian},\ and\ \citenamefont
  {Reinert}}]{Bentmann2017}%
  \BibitemOpen
  \bibfield  {author} {\bibinfo {author} {\bibfnamefont {H.}~\bibnamefont
  {Bentmann}}, \bibinfo {author} {\bibfnamefont {H.}~\bibnamefont {Maa\ss{}}},
  \bibinfo {author} {\bibfnamefont {E.~E.}\ \bibnamefont {Krasovskii}},
  \bibinfo {author} {\bibfnamefont {T.~R.~F.}\ \bibnamefont {Peixoto}},
  \bibinfo {author} {\bibfnamefont {C.}~\bibnamefont {Seibel}}, \bibinfo
  {author} {\bibfnamefont {M.}~\bibnamefont {Leandersson}}, \bibinfo {author}
  {\bibfnamefont {T.}~\bibnamefont {Balasubramanian}}, \ and\ \bibinfo {author}
  {\bibfnamefont {F.}~\bibnamefont {Reinert}},\ }\bibfield  {title} {\enquote
  {\bibinfo {title} {{Strong Linear Dichroism in Spin-Polarized Photoemission
  from Spin-Orbit-Coupled Surface States}},}\ }\href {\doibase
  10.1103/PhysRevLett.119.106401} {\bibfield  {journal} {\bibinfo  {journal}
  {Phys. Rev. Lett.}\ }\textbf {\bibinfo {volume} {119}},\ \bibinfo {pages}
  {106401} (\bibinfo {year} {2017})}\BibitemShut {NoStop}%
\bibitem [{\citenamefont {Okuda}\ \emph {et~al.}(2011)\citenamefont {Okuda},
  \citenamefont {Miyamaoto}, \citenamefont {Miyahara}, \citenamefont {Kuroda},
  \citenamefont {Kimura}, \citenamefont {Namatame},\ and\ \citenamefont
  {Taniguchi}}]{Okuda_RSI_2011}%
  \BibitemOpen
  \bibfield  {author} {\bibinfo {author} {\bibfnamefont {Taichi}\ \bibnamefont
  {Okuda}}, \bibinfo {author} {\bibfnamefont {Koji}\ \bibnamefont {Miyamaoto}},
  \bibinfo {author} {\bibfnamefont {Hirokazu}\ \bibnamefont {Miyahara}},
  \bibinfo {author} {\bibfnamefont {Kenta}\ \bibnamefont {Kuroda}}, \bibinfo
  {author} {\bibfnamefont {Akio}\ \bibnamefont {Kimura}}, \bibinfo {author}
  {\bibfnamefont {Hirofumi}\ \bibnamefont {Namatame}}, \ and\ \bibinfo {author}
  {\bibfnamefont {Masaki}\ \bibnamefont {Taniguchi}},\ }\bibfield  {title}
  {\enquote {\bibinfo {title} {Efficient spin resolved spectroscopy observation
  machine at {Hiroshima Synchrotron Radiation Center}},}\ }\href {\doibase
  10.1063/1.3648102} {\bibfield  {journal} {\bibinfo  {journal} {Rev. Sci.
  Instrum.}\ }\textbf {\bibinfo {volume} {82}},\ \bibinfo {pages} {103302}
  (\bibinfo {year} {2011})}\BibitemShut {NoStop}%
\bibitem [{\citenamefont {Kliemt}\ \emph {et~al.}(2019)\citenamefont {Kliemt},
  \citenamefont {Peters}, \citenamefont {Feldmann}, \citenamefont {Kraiker},
  \citenamefont {Tran}, \citenamefont {Rongstock}, \citenamefont {Hellwig},
  \citenamefont {Witt}, \citenamefont {Bolte},\ and\ \citenamefont
  {Krellner}}]{Kliemt_2019}%
  \BibitemOpen
  \bibfield  {author} {\bibinfo {author} {\bibfnamefont {K.}~\bibnamefont
  {Kliemt}}, \bibinfo {author} {\bibfnamefont {M.}~\bibnamefont {Peters}},
  \bibinfo {author} {\bibfnamefont {F.}~\bibnamefont {Feldmann}}, \bibinfo
  {author} {\bibfnamefont {A.}~\bibnamefont {Kraiker}}, \bibinfo {author}
  {\bibfnamefont {D.-M.}\ \bibnamefont {Tran}}, \bibinfo {author}
  {\bibfnamefont {S.}~\bibnamefont {Rongstock}}, \bibinfo {author}
  {\bibfnamefont {J.}~\bibnamefont {Hellwig}}, \bibinfo {author} {\bibfnamefont
  {S.}~\bibnamefont {Witt}}, \bibinfo {author} {\bibfnamefont {M.}~\bibnamefont
  {Bolte}}, \ and\ \bibinfo {author} {\bibfnamefont {C.}~\bibnamefont
  {Krellner}},\ }\bibfield  {title} {\enquote {\bibinfo {title} {Crystal growth
  of materials with the {ThCr$_2$Si$_2$} structure type},}\ }\href {\doibase
  10.1002/crat.201900116} {\bibfield  {journal} {\bibinfo  {journal} {Crystal
  Research and Technology}\ ,\ \bibinfo {pages} {1900116}} (\bibinfo {year}
  {2019})}\BibitemShut {NoStop}%
\bibitem [{\citenamefont {Koepernik}\ and\ \citenamefont
  {Eschrig}(1999)}]{Koepernik1999}%
  \BibitemOpen
  \bibfield  {author} {\bibinfo {author} {\bibfnamefont {K.}~\bibnamefont
  {Koepernik}}\ and\ \bibinfo {author} {\bibfnamefont {H.}~\bibnamefont
  {Eschrig}},\ }\bibfield  {title} {\enquote {\bibinfo {title} {{Full-potential
  nonorthogonal local-orbital minimum-basis band-structure scheme}},}\ }\href
  {\doibase 10.1103/PhysRevB.59.1743} {\bibfield  {journal} {\bibinfo
  {journal} {Phys. Rev. B}\ }\textbf {\bibinfo {volume} {59}},\ \bibinfo
  {pages} {1743} (\bibinfo {year} {1999})}\BibitemShut {NoStop}%
\bibitem [{Note1()}]{Note1}%
  \BibitemOpen
  \bibinfo {note} {See Supplemental Material for the geometry of SR-ARPES
  experiments, the details of the DFT calculations, the calculations of spin
  photocurrent from magnetic surfaces based on a large-size relativistic
  ${\protect \mathbf k}\cdot {\protect \mathbf p}$ Hamiltonian, and the
  derivation of four- and two-band relativistic ${\protect \mathbf k}\cdot
  {\protect \mathbf p}$ Hamiltonians up to the third and fourth order in
  \protect \textbf {k}, including the additional Refs.~\cite {Leowdin_JCP_1951,
  Schrieffer_PR_1966, Krasovskii_PRB_1997, Krasovskii_PRB_1999}.}\BibitemShut
  {Stop}%
\bibitem [{\citenamefont {Adawi}(1964)}]{Adawi64}%
  \BibitemOpen
  \bibfield  {author} {\bibinfo {author} {\bibfnamefont {I.}~\bibnamefont
  {Adawi}},\ }\bibfield  {title} {\enquote {\bibinfo {title} {Theory of the
  surface photoelectric effect for one and two photons},}\ }\href@noop {}
  {\bibfield  {journal} {\bibinfo  {journal} {Phys. Rev.}\ }\textbf {\bibinfo
  {volume} {134}},\ \bibinfo {pages} {A788--A798} (\bibinfo {year}
  {1964})}\BibitemShut {NoStop}%
\bibitem [{\citenamefont {Krasovskii}\ and\ \citenamefont
  {Schattke}(1999)}]{Krasovskii99}%
  \BibitemOpen
  \bibfield  {author} {\bibinfo {author} {\bibfnamefont {E.~E.}\ \bibnamefont
  {Krasovskii}}\ and\ \bibinfo {author} {\bibfnamefont {W.}~\bibnamefont
  {Schattke}},\ }\bibfield  {title} {\enquote {\bibinfo {title} {Calculation of
  the wave functions for semi-infinite crystals with linear methods of band
  theory},}\ }\href@noop {} {\bibfield  {journal} {\bibinfo  {journal} {Phys.
  Rev. B}\ }\textbf {\bibinfo {volume} {59}},\ \bibinfo {pages}
  {R15609--R15612} (\bibinfo {year} {1999})}\BibitemShut {NoStop}%
\bibitem [{\citenamefont {Quezel}\ \emph {et~al.}(1984)\citenamefont {Quezel},
  \citenamefont {Rossat-Mignod}, \citenamefont {Chevalier}, \citenamefont
  {Lejay},\ and\ \citenamefont {Etourneau}}]{Quezel_TRS_1984}%
  \BibitemOpen
  \bibfield  {author} {\bibinfo {author} {\bibfnamefont {S.}~\bibnamefont
  {Quezel}}, \bibinfo {author} {\bibfnamefont {J.}~\bibnamefont
  {Rossat-Mignod}}, \bibinfo {author} {\bibfnamefont {B.}~\bibnamefont
  {Chevalier}}, \bibinfo {author} {\bibfnamefont {P.}~\bibnamefont {Lejay}}, \
  and\ \bibinfo {author} {\bibfnamefont {J.}~\bibnamefont {Etourneau}},\
  }\bibfield  {title} {\enquote {\bibinfo {title} {Magnetic ordering in
  {TbRh$_2$Si$_2$} and {CeRh$_2$Si$_2$}},}\ }\href {\doibase
  https://doi.org/10.1016/0038-1098(84)90221-7} {\bibfield  {journal} {\bibinfo
   {journal} {Solid State Commun.}\ }\textbf {\bibinfo {volume} {49}},\
  \bibinfo {pages} {685 -- 691} (\bibinfo {year} {1984})}\BibitemShut {NoStop}%
\bibitem [{\citenamefont {Chikina}\ \emph {et~al.}(2014)\citenamefont
  {Chikina}, \citenamefont {H\"{o}ppner}, \citenamefont {Seiro}, \citenamefont
  {Kummer}, \citenamefont {Danzenb\"{a}cher}, \citenamefont {Patil},
  \citenamefont {Generalov}, \citenamefont {G\"{u}ttler}, \citenamefont
  {Kucherenko}, \citenamefont {Chulkov}, \citenamefont {Koroteev},
  \citenamefont {K\"{o}pernik}, \citenamefont {Geibel}, \citenamefont {Shi},
  \citenamefont {Radovic}, \citenamefont {Laubschat},\ and\ \citenamefont
  {Vyalikh}}]{Chick14}%
  \BibitemOpen
  \bibfield  {author} {\bibinfo {author} {\bibfnamefont {A.}~\bibnamefont
  {Chikina}}, \bibinfo {author} {\bibfnamefont {M.}~\bibnamefont
  {H\"{o}ppner}}, \bibinfo {author} {\bibfnamefont {S.}~\bibnamefont {Seiro}},
  \bibinfo {author} {\bibfnamefont {K.}~\bibnamefont {Kummer}}, \bibinfo
  {author} {\bibfnamefont {S.}~\bibnamefont {Danzenb\"{a}cher}}, \bibinfo
  {author} {\bibfnamefont {S.}~\bibnamefont {Patil}}, \bibinfo {author}
  {\bibfnamefont {A.}~\bibnamefont {Generalov}}, \bibinfo {author}
  {\bibfnamefont {M.}~\bibnamefont {G\"{u}ttler}}, \bibinfo {author}
  {\bibfnamefont {Yu.}\ \bibnamefont {Kucherenko}}, \bibinfo {author}
  {\bibfnamefont {E.~V.}\ \bibnamefont {Chulkov}}, \bibinfo {author}
  {\bibfnamefont {Yu.~M.}\ \bibnamefont {Koroteev}}, \bibinfo {author}
  {\bibfnamefont {K.}~\bibnamefont {K\"{o}pernik}}, \bibinfo {author}
  {\bibfnamefont {C.}~\bibnamefont {Geibel}}, \bibinfo {author} {\bibfnamefont
  {M.}~\bibnamefont {Shi}}, \bibinfo {author} {\bibfnamefont {M.}~\bibnamefont
  {Radovic}}, \bibinfo {author} {\bibfnamefont {C.}~\bibnamefont {Laubschat}},
  \ and\ \bibinfo {author} {\bibfnamefont {D.~V.}\ \bibnamefont {Vyalikh}},\
  }\bibfield  {title} {\enquote {\bibinfo {title} {Strong ferromagnetism at the
  surface of an antiferromagnet caused by buried magnetic moments},}\ }\href
  {\doibase 10.1038/ncomms4171} {\bibfield  {journal} {\bibinfo  {journal}
  {Nature Commun.}\ }\textbf {\bibinfo {volume} {5}},\ \bibinfo {pages} {3171}
  (\bibinfo {year} {2014})}\BibitemShut {NoStop}%
\bibitem [{\citenamefont {G\"uttler}\ \emph {et~al.}(2014)\citenamefont
  {G\"uttler}, \citenamefont {Kummer}, \citenamefont {Patil}, \citenamefont
  {H\"oppner}, \citenamefont {Hannaske}, \citenamefont {Danzenb\"acher},
  \citenamefont {Shi}, \citenamefont {Radovic}, \citenamefont {Rienks},
  \citenamefont {Laubschat}, \citenamefont {Geibel},\ and\ \citenamefont
  {Vyalikh}}]{YbCoSi_2}%
  \BibitemOpen
  \bibfield  {author} {\bibinfo {author} {\bibfnamefont {M.}~\bibnamefont
  {G\"uttler}}, \bibinfo {author} {\bibfnamefont {K.}~\bibnamefont {Kummer}},
  \bibinfo {author} {\bibfnamefont {S.}~\bibnamefont {Patil}}, \bibinfo
  {author} {\bibfnamefont {M.}~\bibnamefont {H\"oppner}}, \bibinfo {author}
  {\bibfnamefont {A.}~\bibnamefont {Hannaske}}, \bibinfo {author}
  {\bibfnamefont {S.}~\bibnamefont {Danzenb\"acher}}, \bibinfo {author}
  {\bibfnamefont {M.}~\bibnamefont {Shi}}, \bibinfo {author} {\bibfnamefont
  {M.}~\bibnamefont {Radovic}}, \bibinfo {author} {\bibfnamefont
  {E.}~\bibnamefont {Rienks}}, \bibinfo {author} {\bibfnamefont
  {C.}~\bibnamefont {Laubschat}}, \bibinfo {author} {\bibfnamefont
  {C.}~\bibnamefont {Geibel}}, \ and\ \bibinfo {author} {\bibfnamefont {D.~V.}\
  \bibnamefont {Vyalikh}},\ }\bibfield  {title} {\enquote {\bibinfo {title}
  {{Tracing the localization of $4f$ electrons: Angle-resolved photoemission on
  ${\mathrm{YbCo}}_{2}{\mathrm{Si}}_{2}$, the stable trivalent counterpart of
  the heavy-fermion ${\mathrm{YbRh}}_{2}{\mathrm{Si}}_{2}$}},}\ }\href
  {\doibase 10.1103/PhysRevB.90.195138} {\bibfield  {journal} {\bibinfo
  {journal} {Phys. Rev. B}\ }\textbf {\bibinfo {volume} {90}},\ \bibinfo
  {pages} {195138} (\bibinfo {year} {2014})}\BibitemShut {NoStop}%
\bibitem [{\citenamefont {G\"{u}ttler}\ \emph {et~al.}(2016)\citenamefont
  {G\"{u}ttler}, \citenamefont {Generalov}, \citenamefont {Otrokov},
  \citenamefont {Kummer}, \citenamefont {Kliemt}, \citenamefont {Fedorov},
  \citenamefont {Chikina}, \citenamefont {Danzenb\"{a}cher}, \citenamefont
  {Schulz}, \citenamefont {Chulkov}, \citenamefont {Koroteev}, \citenamefont
  {Caroca-Canales}, \citenamefont {Shi}, \citenamefont {Radovic}, \citenamefont
  {Geibel}, \citenamefont {Laubschat}, \citenamefont {Dudin}, \citenamefont
  {Kim}, \citenamefont {Hoesch}, \citenamefont {Krellner},\ and\ \citenamefont
  {Vyalikh}}]{GdRhSi}%
  \BibitemOpen
  \bibfield  {author} {\bibinfo {author} {\bibfnamefont {M.}~\bibnamefont
  {G\"{u}ttler}}, \bibinfo {author} {\bibfnamefont {A.}~\bibnamefont
  {Generalov}}, \bibinfo {author} {\bibfnamefont {M.~M.}\ \bibnamefont
  {Otrokov}}, \bibinfo {author} {\bibfnamefont {K.}~\bibnamefont {Kummer}},
  \bibinfo {author} {\bibfnamefont {K.}~\bibnamefont {Kliemt}}, \bibinfo
  {author} {\bibfnamefont {A.}~\bibnamefont {Fedorov}}, \bibinfo {author}
  {\bibfnamefont {A.}~\bibnamefont {Chikina}}, \bibinfo {author} {\bibfnamefont
  {S.}~\bibnamefont {Danzenb\"{a}cher}}, \bibinfo {author} {\bibfnamefont
  {S.}~\bibnamefont {Schulz}}, \bibinfo {author} {\bibfnamefont {E.~V.}\
  \bibnamefont {Chulkov}}, \bibinfo {author} {\bibfnamefont {Yu.~M.}\
  \bibnamefont {Koroteev}}, \bibinfo {author} {\bibfnamefont {N.}~\bibnamefont
  {Caroca-Canales}}, \bibinfo {author} {\bibfnamefont {M.}~\bibnamefont {Shi}},
  \bibinfo {author} {\bibfnamefont {M.}~\bibnamefont {Radovic}}, \bibinfo
  {author} {\bibfnamefont {C.}~\bibnamefont {Geibel}}, \bibinfo {author}
  {\bibfnamefont {C.}~\bibnamefont {Laubschat}}, \bibinfo {author}
  {\bibfnamefont {P.}~\bibnamefont {Dudin}}, \bibinfo {author} {\bibfnamefont
  {T.~K.}\ \bibnamefont {Kim}}, \bibinfo {author} {\bibfnamefont
  {M.}~\bibnamefont {Hoesch}}, \bibinfo {author} {\bibfnamefont
  {C.}~\bibnamefont {Krellner}}, \ and\ \bibinfo {author} {\bibfnamefont
  {D.~V.}\ \bibnamefont {Vyalikh}},\ }\bibfield  {title} {\enquote {\bibinfo
  {title} {Robust and tunable itinerant ferromagnetism at the silicon surface
  of the antiferromagnet {GdRh$_2$Si$_2$}},}\ }\href {\doibase
  10.1038/srep24254} {\bibfield  {journal} {\bibinfo  {journal} {Sci. Rep.}\
  }\textbf {\bibinfo {volume} {6}},\ \bibinfo {pages} {24254} (\bibinfo {year}
  {2016})}\BibitemShut {NoStop}%
\bibitem [{\citenamefont {Generalov}\ \emph {et~al.}(2017)\citenamefont
  {Generalov}, \citenamefont {Otrokov}, \citenamefont {Chikina}, \citenamefont
  {Kliemt}, \citenamefont {Kummer}, \citenamefont {H\"{o}ppner}, \citenamefont
  {G\"{u}ttler}, \citenamefont {Seiro}, \citenamefont {Fedorov}, \citenamefont
  {Schulz}, \citenamefont {Danzenb\"{a}cher}, \citenamefont {Chulkov},
  \citenamefont {Geibel}, \citenamefont {Laubschat}, \citenamefont {Dudin},
  \citenamefont {Hoesch}, \citenamefont {Kim}, \citenamefont {Radovic},
  \citenamefont {Shi}, \citenamefont {Plumb}, \citenamefont {Krellner},\ and\
  \citenamefont {Vyalikh}}]{HoRhSi}%
  \BibitemOpen
  \bibfield  {author} {\bibinfo {author} {\bibfnamefont {Alexander}\
  \bibnamefont {Generalov}}, \bibinfo {author} {\bibfnamefont {Mikhail~M.}\
  \bibnamefont {Otrokov}}, \bibinfo {author} {\bibfnamefont {Alla}\
  \bibnamefont {Chikina}}, \bibinfo {author} {\bibfnamefont {Kristin}\
  \bibnamefont {Kliemt}}, \bibinfo {author} {\bibfnamefont {Kurt}\ \bibnamefont
  {Kummer}}, \bibinfo {author} {\bibfnamefont {Marc}\ \bibnamefont
  {H\"{o}ppner}}, \bibinfo {author} {\bibfnamefont {Monika}\ \bibnamefont
  {G\"{u}ttler}}, \bibinfo {author} {\bibfnamefont {Silvia}\ \bibnamefont
  {Seiro}}, \bibinfo {author} {\bibfnamefont {Alexander}\ \bibnamefont
  {Fedorov}}, \bibinfo {author} {\bibfnamefont {Susanne}\ \bibnamefont
  {Schulz}}, \bibinfo {author} {\bibfnamefont {Steffen}\ \bibnamefont
  {Danzenb\"{a}cher}}, \bibinfo {author} {\bibfnamefont {Evgueni~V.}\
  \bibnamefont {Chulkov}}, \bibinfo {author} {\bibfnamefont {Christoph}\
  \bibnamefont {Geibel}}, \bibinfo {author} {\bibfnamefont {Clemens}\
  \bibnamefont {Laubschat}}, \bibinfo {author} {\bibfnamefont {Pavel}\
  \bibnamefont {Dudin}}, \bibinfo {author} {\bibfnamefont {Moritz}\
  \bibnamefont {Hoesch}}, \bibinfo {author} {\bibfnamefont {Timur}\
  \bibnamefont {Kim}}, \bibinfo {author} {\bibfnamefont {Milan}\ \bibnamefont
  {Radovic}}, \bibinfo {author} {\bibfnamefont {Ming}\ \bibnamefont {Shi}},
  \bibinfo {author} {\bibfnamefont {Nicholas~C.}\ \bibnamefont {Plumb}},
  \bibinfo {author} {\bibfnamefont {Cornelius}\ \bibnamefont {Krellner}}, \
  and\ \bibinfo {author} {\bibfnamefont {Denis~V.}\ \bibnamefont {Vyalikh}},\
  }\bibfield  {title} {\enquote {\bibinfo {title} {Spin orientation of
  two-dimensional electrons driven by temperature-tunable competition of
  spin{\textendash}orbit and exchange{\textendash}magnetic interactions},}\
  }\href {\doibase 10.1021/acs.nanolett.6b04036} {\bibfield  {journal}
  {\bibinfo  {journal} {Nano Lett.}\ }\textbf {\bibinfo {volume} {17}},\
  \bibinfo {pages} {811--820} (\bibinfo {year} {2017})}\BibitemShut {NoStop}%
\bibitem [{\citenamefont {Winkler}(2000)}]{Winkler_PRB_2000}%
  \BibitemOpen
  \bibfield  {author} {\bibinfo {author} {\bibfnamefont {R.}~\bibnamefont
  {Winkler}},\ }\bibfield  {title} {\enquote {\bibinfo {title} {Rashba spin
  splitting in two-dimensional electron and hole systems},}\ }\href {\doibase
  10.1103/PhysRevB.62.4245} {\bibfield  {journal} {\bibinfo  {journal} {Phys.
  Rev. B}\ }\textbf {\bibinfo {volume} {62}},\ \bibinfo {pages} {4245--4248}
  (\bibinfo {year} {2000})}\BibitemShut {NoStop}%
\bibitem [{\citenamefont {Winkler}\ \emph {et~al.}(2008)\citenamefont
  {Winkler}, \citenamefont {Culcer}, \citenamefont {Papadakis}, \citenamefont
  {Habib},\ and\ \citenamefont {Shayegan}}]{Winkler_SST_2008}%
  \BibitemOpen
  \bibfield  {author} {\bibinfo {author} {\bibfnamefont {R.}~\bibnamefont
  {Winkler}}, \bibinfo {author} {\bibfnamefont {Dimitrie}\ \bibnamefont
  {Culcer}}, \bibinfo {author} {\bibfnamefont {S.~J.}\ \bibnamefont
  {Papadakis}}, \bibinfo {author} {\bibfnamefont {B.}~\bibnamefont {Habib}}, \
  and\ \bibinfo {author} {\bibfnamefont {M.}~\bibnamefont {Shayegan}},\
  }\bibfield  {title} {\enquote {\bibinfo {title} {Spin orientation of holes in
  quantum wells},}\ }\href {\doibase 10.1088/0268-1242/23/11/114017} {\bibfield
   {journal} {\bibinfo  {journal} {Semicond. Sci. Technol.}\ }\textbf {\bibinfo
  {volume} {23}},\ \bibinfo {pages} {114017} (\bibinfo {year}
  {2008})}\BibitemShut {NoStop}%
\bibitem [{\citenamefont {Schliemann}\ and\ \citenamefont
  {Loss}(2005)}]{Schliemann_PRB_2005}%
  \BibitemOpen
  \bibfield  {author} {\bibinfo {author} {\bibfnamefont {John}\ \bibnamefont
  {Schliemann}}\ and\ \bibinfo {author} {\bibfnamefont {Daniel}\ \bibnamefont
  {Loss}},\ }\bibfield  {title} {\enquote {\bibinfo {title} {{Spin-Hall
  transport of heavy holes in III-V semiconductor quantum wells}},}\ }\href
  {\doibase 10.1103/PhysRevB.71.085308} {\bibfield  {journal} {\bibinfo
  {journal} {Phys. Rev. B}\ }\textbf {\bibinfo {volume} {71}},\ \bibinfo
  {pages} {085308} (\bibinfo {year} {2005})}\BibitemShut {NoStop}%
\bibitem [{\citenamefont {Nechaev}\ and\ \citenamefont
  {Krasovskii}(2016)}]{Nechaev_PRBR_2016}%
  \BibitemOpen
  \bibfield  {author} {\bibinfo {author} {\bibfnamefont {I.~A.}\ \bibnamefont
  {Nechaev}}\ and\ \bibinfo {author} {\bibfnamefont {E.~E.}\ \bibnamefont
  {Krasovskii}},\ }\bibfield  {title} {\enquote {\bibinfo {title} {Relativistic
  $\mathrm{k}\ifmmode\cdot\else\textperiodcentered\fi{}\mathrm{p}$ hamiltonians
  for centrosymmetric topological insulators from \textit{ab initio} wave
  functions},}\ }\href {\doibase 10.1103/PhysRevB.94.201410} {\bibfield
  {journal} {\bibinfo  {journal} {Phys. Rev. B}\ }\textbf {\bibinfo {volume}
  {94}},\ \bibinfo {pages} {201410} (\bibinfo {year} {2016})}\BibitemShut
  {NoStop}%
\bibitem [{\citenamefont {L\"{o}wdin}(1951)}]{Leowdin_JCP_1951}%
  \BibitemOpen
  \bibfield  {author} {\bibinfo {author} {\bibfnamefont {Per-Olov}\
  \bibnamefont {L\"{o}wdin}},\ }\bibfield  {title} {\enquote {\bibinfo {title}
  {{A Note on the Quantum-Mechanical Perturbation Theory}},}\ }\href {\doibase
  10.1063/1.1748067} {\bibfield  {journal} {\bibinfo  {journal} {J. Chem.
  Phys.}\ }\textbf {\bibinfo {volume} {19}},\ \bibinfo {pages} {1396--1401}
  (\bibinfo {year} {1951})}\BibitemShut {NoStop}%
\bibitem [{\citenamefont {Schrieffer}\ and\ \citenamefont
  {Wolff}(1966)}]{Schrieffer_PR_1966}%
  \BibitemOpen
  \bibfield  {author} {\bibinfo {author} {\bibfnamefont {J.~R.}\ \bibnamefont
  {Schrieffer}}\ and\ \bibinfo {author} {\bibfnamefont {P.~A.}\ \bibnamefont
  {Wolff}},\ }\bibfield  {title} {\enquote {\bibinfo {title} {{Relation between
  the Anderson and Kondo Hamiltonians}},}\ }\href {\doibase
  10.1103/PhysRev.149.491} {\bibfield  {journal} {\bibinfo  {journal} {Phys.
  Rev.}\ }\textbf {\bibinfo {volume} {149}},\ \bibinfo {pages} {491--492}
  (\bibinfo {year} {1966})}\BibitemShut {NoStop}%
\bibitem [{\citenamefont {Krasovskii}(1997)}]{Krasovskii_PRB_1997}%
  \BibitemOpen
  \bibfield  {author} {\bibinfo {author} {\bibfnamefont {E.~E.}\ \bibnamefont
  {Krasovskii}},\ }\bibfield  {title} {\enquote {\bibinfo {title} {Accuracy and
  convergence properties of the extended linear augmented-plane-wave method},}\
  }\href {\doibase 10.1103/PhysRevB.56.12866} {\bibfield  {journal} {\bibinfo
  {journal} {Phys. Rev. B}\ }\textbf {\bibinfo {volume} {56}},\ \bibinfo
  {pages} {12866--12873} (\bibinfo {year} {1997})}\BibitemShut {NoStop}%
\bibitem [{\citenamefont {Krasovskii}\ \emph {et~al.}(1999)\citenamefont
  {Krasovskii}, \citenamefont {Starrost},\ and\ \citenamefont
  {Schattke}}]{Krasovskii_PRB_1999}%
  \BibitemOpen
  \bibfield  {author} {\bibinfo {author} {\bibfnamefont {E.~E.}\ \bibnamefont
  {Krasovskii}}, \bibinfo {author} {\bibfnamefont {F.}~\bibnamefont
  {Starrost}}, \ and\ \bibinfo {author} {\bibfnamefont {W.}~\bibnamefont
  {Schattke}},\ }\bibfield  {title} {\enquote {\bibinfo {title} {Augmented
  fourier components method for constructing the crystal potential in
  self-consistent band-structure calculations},}\ }\href {\doibase
  10.1103/PhysRevB.59.10504} {\bibfield  {journal} {\bibinfo  {journal} {Phys.
  Rev. B}\ }\textbf {\bibinfo {volume} {59}},\ \bibinfo {pages} {10504--10511}
  (\bibinfo {year} {1999})}\BibitemShut {NoStop}%
\end{thebibliography}
\end{document}